\documentclass[prd,showpacs,preprintnumbers,amsmath,amssymb,floatfix]{revtex4}

\usepackage{graphicx}
\usepackage{dcolumn}
\usepackage{bm}
\usepackage{amssymb}
\usepackage{epsfig}
\usepackage{color}

\newcommand{\ba}{\begin{eqnarray}}
\newcommand{\ea}{\end{eqnarray}}
\newcommand{\be}{\begin{equation}}
\newcommand{\ee}{\end{equation}}
\newcommand{\bdisplay}{\begin{displaymath}}
\newcommand{\edisplay}{\end{displaymath}}

\newcommand{\eq}[1]{Eq.\,(\ref{#1})}

\begin{document}

\title{Implications of a Froissart bound saturation of $\gamma^*$-$p$ deep inelastic scattering. Part I.  Quark distributions at ultra small $x$.}
\author{Martin~M.~Block}
\email{mblock@northwestern.edu}
\affiliation{Department of Physics and Astronomy, Northwestern University,
Evanston, IL 60208}
\author{Loyal Durand}
\email{ldurand@hep.wisc.edu}
\altaffiliation{Mailing address: 415 Pearl Ct., Aspen, CO 81611}
\affiliation{Department of Physics, University of Wisconsin, Madison, WI 53706}
\author{Phuoc Ha}
\email{pdha@towson.edu}
\affiliation{Department of Physics, Astronomy and Geosciences, Towson University, Towson, MD 21252}
\author{Douglas W. McKay}
\email{dmckay@ku.edu}
\affiliation{Department of Physics and Astronomy, University of Kansas, Lawrence, KS 66045}
%\begin{document}
\begin{abstract}
We argue that the  deep inelastic structure function $F_2^{\gamma p}(x, Q^2)$, regarded as a cross section for virtual $\gamma^*p$ scattering,  is hadronic in nature. This implies that its growth is limited by the Froissart bound at high hadronic energies, giving a $\ln^2 (1/x)$ bound on $F_2^{\gamma p}$ as Bjorken $x\rightarrow 0$. The same bound holds for the individual quark distributions.  In earlier work, we obtained a very accurate  global fit to the combined HERA data on $F_2^{\gamma p}$ using a fit function which respects the Froissart bound at small $x$, and is equivalent in its $x$ dependence to the function used successfully to describe all high energy hadronic cross sections, including $\gamma p$ scattering. We  extrapolate that fit by a factor of $\lesssim$3  beyond the HERA region in the natural variable $\ln(1/x)$ to the values of $x$ down to $x=10^{-14}$  and use the results to derive the quark distributions needed for the reliable calculation of neutrino cross sections  at energies up to $E_\nu=10^{17}$ GeV. These distributions do not satisfy the Feynman ``wee parton" assumption, that they all converge toward a common distribution $xq(x,Q^2)$ at small $x$ and large $Q^2$. This was used in some past calculations to express the dominant neutrino structure function   $F_2^{\nu(\bar{\nu})}$ directly in terms of  $F_2^{\gamma p}$. We show that the correct distributions nevertheless give results for  $F_2^{\nu(\bar{\nu})}$ which differ only slightly from those obtained assuming that the wee parton limit holds. In two Appendices, we develop simple analytic results for the effects of QCD evolution and operator-product corrections on the distribution functions at small $x$, and show that these effects amount mainly to shifting the values of $\ln(1/x)$ in the initial distributions.

\end{abstract}

\date{\today}

\maketitle

%%%%%%%%%%%%% SEC. INTRODUCTION %%%%%%%%%%

\section{Introduction \label{sec:Introduction}}

%%%%%%%%%%%%%%%%%%%%%%%%%%%%%%%%%%%

The experimental program at HERA, the electron-proton collider at DESY, probed deep inelastic scattering (DIS) at small values of the Bjorken variable $x$, given in terms of the proton momentum $p$ and the electron momentum transfer $q$ in the scattering by $x\approx Q^2/2p\cdot q$. The measurements covered  the range $10^{-6}$ to $10^{-1}$  in $x$,  with a corresponding range from 0.1 GeV$^2$ to 5000 GeV$^2$ for the photon virtuality $Q^2=-q^2$ .  The results of the extensive measurements  by the H1 and ZEUS detector groups show that the structure function $F_2^{\gamma p}(x, Q^2)$ rises rapidly as $x$ decreases with $Q^2$ fixed.

 It has been argued in a series of papers over the past few years  \cite{bbt1,bbt2, bbmt, bhm}  that the reduced cross section in $ep$ (or $\gamma^*p$) DIS, basically the structure function $F_2^{\gamma p}(x,Q^2)$, is hadronic in nature and satisfies a saturated Froissart bound, implying that $F_2^{\gamma p}(x,Q^2) \rightarrow {\rm constant}\times \ln(1/x)^2$ for $x\rightarrow0$ with $Q^2$ fixed.  The basic argument, reviewed and extended here,  is that the structure function $F_2^{\gamma p}$ is determined by the total cross section of the off-shell photon $\gamma^*$ on the proton at a $\gamma^*p$ center-of-mass energy squared $\hat{W}^2=\hat s$, where $\hat s=(p+q)^2$ is the usual Mandelstam variable, and is thus  subject through analyticity and unitarity constraints to the saturated Froissart bound on total hadronic cross sections  $\sigma(\hat s )\rightarrow\sigma_0\log^2 \hat s$ as  $ \hat s\rightarrow\infty$.

 The picture is compelling in light of its success in describing hadron-hadron and photon-hadron total cross sections over many orders of magnitude in $s$ with the same basic functional form \cite{blockrev}. Moreover, the \emph{predictions} for the proton-proton and proton-air cross sections at the LHC \cite{LHCtot1,LHCtot2,LHCtot3} and the Pierre Auger Observatory \cite{POAp-air}, respectively, obtained by extrapolating that form to much higher $s$ are confirmed by these new high energy experiments \cite{mbair,blockhalzen,blockhalzen2}.

In this paper, we investigate the implications of this bounded behavior for $F_2^{\gamma p}$ for  the ultra-small $x$,  large $Q^2$  limit of the quark distributions in the proton using our recent Froissart-bounded fit to the combined HERA data. This fit can potentially be checked at the proposed Large Hadron electron Collider  (LHeC) \cite{LHeC} over ranges in $x$ and $Q^2$ larger by factors of $\sim20$ than those accessible at HERA.

We show that the individual quark distributions can be derived to good accuracy directly from $F_2^{\gamma p}$, and present the results obtained using the extrapolation of our  fit to ultra-small $x$.  The extrapolation of $F_2^{\gamma p}$ and the resulting quark distributions should be reliable: the fit function becomes a simple quadratic in  the natural variable $v=\ln(1/x)$ with well-determined coefficients for $x$ small or $v$ large, and the extrapolation necessary to reach $x=10^{-14}$ ($v=32.2$) involves only a factor of $\lesssim$3 increase in $v$ from the upper values attained in the  HERA region.

 In Part II of this work, the companion paper  to this  \cite{bdhmneutrino}, we use the quark distributions derived here to calculate UHE  charged- and neutral current neutrino-nucleon cross sections $\sigma_{CC}^{\nu}(E_{\nu})$ and $\sigma_{NC}^{\nu}(E_{\nu})$ for neutrino energies up to  $E_{\nu}=10^{17}$ GeV, currently the highest energies at which there are experimental bounds on cosmic neutrino fluxes \cite{forte, glue}. These calculations require quark distributions at large $Q^2$ ($Q^2\gtrsim 10^4$) and small $x$, down to $x\sim 10^{-14}$. The results presented here are motivated by that need, and  by the fact that  the  measurement of UHE neutrino cross sections would provide a test of hadronic dynamics at energies not otherwise accessible \cite{bdhmneutrino}.

In some earlier calculations \cite{bbmt,bhm} of ultra high energy (UHE) neutrino-nucleon cross sections, it was assumed that the quark distributions could be treated in Feynman's ``wee parton'' limit in which the  individual quark distributions all converge to a common  quark distribution $xq(x,Q^2)$   at large $Q^2$ and small $x$. This allowed the replacement of individual quark distributions in the neutrino cross sections by a common distribution $xq$. Ignoring QCD corrections, this is given in leading-order  (LO) in terms of $F_2^{\gamma p}$ by  relation  $xq=F_2^{\gamma p}/\sum_ie_i^2$, where the $e_i$ are the quark charges and the sum runs over the active quarks and antiquarks. This relation was  used in \cite{bbmt,bhm} to predict neutrino cross sections at ultra high energies in terms of $F_2^{\gamma p}$.

We show here that the wee parton condition is {\em not} satisfied by the quark distributions determined directly from our fit to $F_2^{\gamma p}$.  However, the relations between $F_2^{\gamma p}$ and the corresponding charged- and neutral-current structure functions $F_2^{\nu(\bar{\nu})}$ and $F0_2^{\nu(\bar{\nu})}$ in neutrino and antineutrino scattering which were derived using the wee parton assumption continue to hold to high accuracy.

The present paper is organized as follows.  In Sec. \ref{extrapolations}, we develop our arguments with respect to the relevance of the Froissart bound and its consequences for the form used in  our parameterization of the small-$x$ HERA data for $F_2^{\gamma p}(x,Q^2)$. We then summarize the results of our fit  \cite{bdhmapp}  to the combined HERA data \cite{HERAcombined}, and list the parameters, their errors and the significance of the fit. In Sec.\ \ref{subsec:relations_for_distributions}, we discuss the derivation of the {\em individual} small $x$ quark distributions from our analytic fit to $F_2^{\gamma p}(x,Q^2)$. This requires information on the singlet quark distribution $F_s(x,Q^2)$ which can be expressed in terms of a ``bare'' structure function $F_{20}^{\gamma p}$ and a set of non-singlet quark distributions.

In two Appendices,  we  develop simple new analytic results for the effects of QCD evolution at small $x$ on the non-singlet distributions, and  show how $F_{20}^{\gamma p}$ can be related analytically to our fit to $F_2^{\gamma p}$. Our results for quark distributions at ultra-small $x$ are given in Sec.\ \ref{subsec:results_for_quarks}. We discuss their implications with respect to the wee parton picture and neutrino cross sections in Sec.\ \ref{sec:wee_partons}, where we show that $F_2^{\nu(\bar{\nu})}$, the dominant neutrino structure function, can be expressed directly in terms of $F_2^{\gamma p}$ to good approximation despite the failure of the wee parton limit used in earlier discussions of this connection  \cite{bbmt,bhm}.
We summarize and draw conclusions in Sec. \ref{sec:conclusions}.

%%%%%%%%%%%%%%%%%%%%%%%%%%%%%%%%%%%
%%%%%%%%%% SEC. EXTRAPOLATION OF F2 %%%%%%%%%

\section{Extrapolation of $F_2^{\gamma p}(x,Q^2)$ to ultra-small $x$ \label{extrapolations}}

\subsection{Background\label{subset:background}}

%%%%%%%%%%%%%%%%%%%%%%%%%%%%%%%%%%%

As we emphasized in the Introduction, for the energies $E_\nu$  of interest for UHE neutrino cross sections, we must know quark distributions at values of $x$ many orders of magnitude below the range where they have been derived from HERA measurements of $F_2^{\gamma p}(x,Q^2)$. In particular, to reach the highest energy range covered by neutrino telescope searches \cite{forte, glue}, $E_\nu\sim 10^{17}$ GeV, requires an extrapolation of eight orders of magnitude  below the lowest values $x\sim 10^{-5}$--$10^{-6}$ encountered at HERA. While this involves only a factor of $\lesssim 3$ increase in the maximum value of the natural variable $v\equiv\ln(1/x)$,
it is still essential that the form used to extrapolate $F_2^{\gamma p}(x,Q^2)$ be consistent  both with the asymptotic limiting behavior expected theoretically, and with the present data. We stress that, if a fit to the data indicates that the measured structure function is already consistent with the limiting asymptotic form, the extrapolation may be expected to be robust; our approach, which we summarize here, has this feature.

The structure function $F_2^{\gamma p}(x,Q^2)$ is equal, up to known (kinematic) factors, to
 the total  cross section  for virtual photon-nucleon $(\gamma^*p)$ scattering, and  contains all of the strong interaction dynamics in the process, a point made clearly in Ref.\ \cite{bbt1}.   It is just the extension of real $\gamma p$  scattering with photon 4-momentum squared $q^2=0$, to virtual $\gamma^*p$ scattering with a virtual photon 4-momentum squared $q^2=-Q^2<0$. We denote the Mandelstam variables for $\gamma^*p$ scattering by $ \hat{s},\,\hat{t},$ and $\hat{u}$.

There is no theoretical obstacle to the continuation from $q^2=0$ to an off-shell $q^2=-Q^2<0$. The total cross section for $\gamma^*p$ scattering is proportional to the virtual forward Compton scattering amplitude for zero momentum transfer to the nucleon, $\hat{t} = (p-p')^2 =0$. A complete all-orders analysis of the latter  in perturbation theory  \cite{ashok} shows that it is real analytic in the Mandelstam variable $ \hat{s}=(p+q)^2$ for the scattering of a virtual photon from a nucleon for $Q^2 > - m_\pi^2$, with the usual normal thresholds in $ \hat{s}$ and $\hat{u}$. Analyticity in $\hat{t}$ can also be established for $Q^2>0$ for the leading perturbative diagrams, and presumably holds in general. Given these results, the arguments of Martin \cite{martin1,martin2,martin3} establish the Froissart bound in $\hat{s}$ \cite{froissart} for the $\gamma^*p$  cross section, hence $F_2^{\gamma p}$.

 The usefulness of  off-shell continuation in masses is shown experimentally by the phenomenological success of the vector meson dominance picture of electromagnetic current matrix elements between hadronic states  \cite{sakurai1,sakurai2,schildknecht}.
In the energy domain, extensive analyses of experimental data on high energy hadronic and photo-production cross sections dramatically demonstrate the early appearance of the $\ln^2 s$  Froissart-like behavior as the Mandelstam variable $s$ for hadron-hadron or photon-hadron scattering increases \cite{blockrev,mbair,blockhalzen}. This can be understood in terms of QCD processes at the quark-gluon level \cite{Snowmass84,Margolis,DurandPi,HDGPS}, with the hadron becoming essentially a ``black disk'' of gluons and quarks when seen at very high energies \cite{blockhalzen2}.  Since deep inelastic  $\gamma^*p$  scattering is smoothly connected to $\gamma p$ scattering by continuation in $Q^2$ \cite{ashok}, it is natural to assume the $\ln^2  s$ Froissart behavior of the photo-production cross section will also appear in DIS  for high $\gamma^* p$ energies with the substitution  $s\rightarrow  \hat{s}$. In fact, detailed perturbative arguments  \cite{GLRsmxqcd} indicate that unitarity begins to be violated at remarkably small values of $v=\ln(1/x)$, e.g.\  for $v\agt 3$ at $Q^2=10^4$ GeV$^2$, in the usual description of the QCD evolution of $F_2^{\gamma p}$ through the   Dokshitzer-Gribov-Lipitov-Altarelli-Parisi (DGLAP) equations \cite{dglap1,dglap2,dglap3}, suggesting the onset of non-perturbative behavior.

The role of Mandelstam $s$ in hadron-hadron scattering is played in $\gamma^*p$ scattering by $ \hat{s}=\hat{W}^2$, the final state hadronic energy squared:
\be
 \hat{s} \equiv \hat{W}^2=(q+p)^2 = \frac{Q^2}{x}(1-x)+m^2 \rightarrow \frac{Q^2}{x},  \quad x \rightarrow 0.
 \label{Wsq}
 \ee
For  $Q^2$ fixed and large relative to the square of the nucleon mass, i.e., $Q^2\gg m^2$, Eq.\ (\ref{Wsq}) shows that a saturated  $\ln^2{ \hat{s}}$ bound on the $\gamma^*p$ cross section translates into a $\ln^2(1/x)=v^2$ bound on the small $x$ (large $v$) behavior of $F_2^{\gamma p}$.

% provides physical evidence that the required off-shell continuation from photo-production to DIS electron-nucleon scattering is correct and relevant \cite{sakurai1,sakurai2,schildknecht}.

%%%%%%%%%% SUBSECTION FROISSART-BOUNDED FIT %%%%%%%%%%%%

\subsection{Froissart-bounded fit to $F_2^{\gamma p}$ \label{subsc:fit_and_neutinos}}

%%%%%%%%%%%%%%%%%%%%%%%%%%%%%%%%

In their analysis of the early ZEUS data on  $F_2^{\gamma p}(x,Q^2)$ \cite{ZEUS1,ZEUS2}, Berger, Block, and Tan \cite{bbt1} assumed that the $\gamma^*p$ cross section should show Froissart-like behavior in $1/x$. The success of their model amply supports this assumption. In a subsequent paper \cite{bbt2}, those authors refined their  saturated ``Froissart'' parameterization and obtained an excellent global fit to both the $x$ and $Q^2$ dependence of the ZEUS data, with  6 free parameters describing hundreds of points of data. This fit was later used along with the Feynman wee parton picture to predict the UHE $\nu N$ cross sections \cite{bbmt}.

Releasing one more parameter, the present authors  \cite{bhm,bdhmapp} fit the joint ZEUS \cite{ZEUS1,ZEUS2} and H1 \cite{H1} determinations of the $e^{\pm}p$ DIS cross sections as combined by those groups \cite{HERAcombined}, a combination that resolved some of the tension between previous individual ZEUS and H1 analyses.  This fit, summarized below, provides very accurate values of $F_2^{\gamma p}(x,Q^2)$ over a large region of the $x$-$Q^2$ plane that includes some 335 data points.

The global fit function used in \cite{bbt2} and \cite{bdhmapp}, which ensures that the saturated Froissart $\ln^2(1/x)$  behavior dominates at small $x$, takes  the form
 \begin{eqnarray}
 F_2^{\gamma p}(x,Q^2) &=& (1-x)\left[\frac{F_P}{1-x_P}+A(Q^2)\ln\left(\frac{x_P}{x}\frac{1-x}{1-x_P}\right)  \right. \nonumber \\
                    &  & \left. +B(Q^2)\ln^2\left(\frac{x_P}{x}\frac{1-x}{1-x_P}\right)\right],
 \label{eqnF2smx}
 \end{eqnarray}
 where
 \begin{eqnarray}
 A(Q^2) &=& a_0+a_1\ln Q^2+a_2\ln^2Q^2, \nonumber \\
 B(Q^2) &=& b_0+b_1\ln Q^2+b_2\ln^2Q^2.
 \label{eqnABsmx}
 \end{eqnarray}
 As is evident from \eq{Wsq}, this form is equivalent at small $x$ to the quadratic expression in $\ln s$ familiar in fits to hadronic data \cite{blockrev}, with the $Q^2$ dependence rearranged and extended.

 At small $x$ or large $v=\ln(1/x)$, the expression in \eq{eqnF2smx} becomes a quadratic polynomial in $v$ with
 \be
 \label{F2hat_v}
\hat{ F}_2^{\gamma p}(v,Q^2)\equiv F_2^{\gamma p}(e^{-v},Q^2) \rightarrow \hat{C}_{0f}(Q^2)+\hat{C}_{1f}(Q^2)v+\hat{C}_{2f}(Q^2)v^2 +O(e^{-v}).
\ee
The coefficients $\hat{C}_i$  are again quadratics in $\ln Q^2$,
\ba
\label{C0}
\hat{C}_{0f}(Q^2) &=& F_P/(1-x_P)+A(Q^2)v_0+B(Q^2)v_0^2, \\
\label{C1}
\hat{C}_{1f}(Q^2) &=& A(Q^2)+2B(Q^2)v_0, \\
\label{C2}
\hat{C}_{2f}(Q^2) &=& B(Q^2),
\ea
where $v_0=\ln[x_P/(1-x_P)]$. We will use this quadratic structure in $v$ repeatedly in the analysis below. As we will see in the Appendices, the neglect of the terms of order $e^{-v}$ in \eq{F2hat_v}, important for $v\sim 0$, will not affect our results at large $v$.

The procedure used in fitting the combined HERA data is described in references \cite{sieve} and \cite{bdhmapp}.
The parameter $x_P$ was fixed at the value 0.11; the HERA data are sparse for larger $x$. $F_P$, the value of $F_2^{\gamma p}$ at $x_P$, and the other 6 fitting parameters are listed in Table \ref{table:results} together with their errors.  Also shown are the renormalized minimized $\chi^2$ value \cite{sieve}, the number of degrees of freedom and the renormalized $\chi^2$ per degree of freedom for our new analytic form for the combined ZEUS and H1 results \cite{HERAcombined}.
\begin{table}[ht]                   % Use "table" environment, but also
                 % use  "tabular" environment below.
%
\begin{center}
\def\arraystretch{1.2}            % Make the space between rows in the Table,
                  % 1.2 x bigger than the default spacing.
%\vspace{.5in}
     \caption{\label{fitted}\protect\small Results of a 7-parameter fit to the
HERA combined data for $F_2^{\gamma p}(x,Q^2)$ for $0.85 \le
Q^2\le 3000$ GeV$^2$ and $x\le 0.1$. The $\chi^2_{\rm min}$ is renormalized  by the factor $\mathcal{R}=1.1$ to take into account the effects of the cut at $\Delta \chi^2_{i, \rm max}$ = 6 introduced by the sieve algorithm used in the fit \cite{sieve}. \label{table:results}}
\begin{tabular}[b]{|l||l||c||}
   % \multicolumn{1}{c}{Parameters}&{Values}\\
    \cline{1-2}
Parameters&Values\\
\hline
      $a_0$&$-8.471\times 10^{-2}\pm 2.62\times 10^{-3}$ \\
      $a_1$&$\phantom{-}4.190\times 10^{-2}\pm 1.56\times 10^{-3}$\\
      $a_2$&$-3.976\times 10^{-3}\pm 2.13\times 10^{-4}$\\
\hline
    $b_0$ &$\phantom{-}1.292\times 10^{-2}\pm 3.62\times 10^{-4}$\\
      $b_1$&$\phantom{-}2.473\times 10^{-4}\pm 2.46\times 10^{-4}$\\
      $b_2$&$\phantom{-}1.642\times 10^{-3}\pm 5.52\times 10^{-5}$ \\
\hline
$F_P$&\phantom{---}$0.413\pm0.003$\\
    \cline{1-2}
        \hline
    \hline
    $\chi^2_{\rm min}$&\phantom{---}352.8\\
    $\mathcal{R}\times\chi^2_{\rm min}$&\phantom{---}391.4\\

    d.o.f.&\phantom{---}335\\
    $\mathcal{R}\times\chi^2_{\rm min}/$d.o.f.&\phantom{---}1.17\\
\hline
\end{tabular}
     %\vspace{1in} \\
\end{center}
\label{tabF2smx}
\end{table}
\def\arraystretch{1}

The high quality of our Froissart-bounded fit to data that range at the limits over $\sim 5$ orders of magnitude in $x$ and $\sim 3$ orders of magnitude in $Q^2$ lends strong support to the proposal that the cross section for nucleon scattering with off-shell photons obeys the saturated Froissart bound in the $\gamma^*p$ Mandelstam variable $\hat s=\hat{W}^2$. The errors in the parameters $a_i$ and $b_i$ are typically a few percent except for $b_1$ which is not well determined, with a size and error  of the order of the errors in the other parameters. It should be emphasized (see Table \ref {table:results}) that the data used in this QCD fit start at $Q^2=0.85$ GeV$^2$ and $x\sim 10^{-6}$, in a region with $Q^2$ so small that perturbative QCD is {\em not} expected to be valid. Since the fit depends linearly on the parameters, the errors propagate linearly, and  the correlated percentage errors in the extrapolation of our fit to  ultra-small $x$  are quite small.

%%%%%%%%%%%%%%%%%%%%%%%%%%%%%%%%%%%%%
%%%%%%% SEC. DERIVATION OF QUARK DISTRIBUTIONS %%%%%

{\section{  DERIVATION OF  QUARK DISTRIBUTIONS AT VERY LOW $x$ FROM $F_2^{\gamma p }$ \label{sec:derivations}}

%%%%%%% SUBSEC. RELATIONS FOR QUARKS %%%%%%%%%%%

\subsection{ Relations for the quark distributions \label{subsec:relations_for_distributions}}

%%%%%%%%%%%%%%%%%%%%%%%%%%

We start by introducing the standard  non-singlet (NS) quark distributions \cite{esw}
\ba
\label{V_i}
V_i &=& x( q_i-\bar{q}_i),\quad i=1,2,\ldots,\\
\label{T3}
T_3 &=&  x( u+\bar{u}-d-\bar{d}), \\
\label{T8}
T_8 &=&  x( u+\bar{u}+d+\bar{d}-2s-2\bar{s}), \\
\label{T15}
T_{15} &=&  x( u+\bar{u}+d+\bar{d}+s+\bar{s}-3c-3\bar{c}), \\
\label{T24}
T_{24} &=&  x( u+\bar{u}+d+\bar{d}+s+\bar{s}+c+\bar{c}-4b-4\bar{b}),
\ea
and the singlet distribution
\be
F_s =  x( u+\bar{u}+d+\bar{d}+s+\bar{s}+c+\bar{c}+b+\bar{b}+\cdots),
\ee
where the quark distributions are all defined at a given order in perturbative QCD.

We will be concerned mainly with very small $x$. We will take $s=\bar{s}$, $c=\bar{c}$, and $b=\bar{b}$, and will neglect the   small differences between the $\bar{u}$ and $\bar{d}$ quarks seen at large $x$. The effects of the valence quark distributions $u_v=V_1$ and $d_v=V_2$ are also quite small at small $x$, and we will take $d_v=(1/2)u_v$, with $u_v\equiv U$, a reasonable approximation, while $T_3\rightarrow (1/2)U$.  $F_s$ is then related to the $\gamma^*p$ structure function  for different numbers $n_f$ of active quarks as
\ba
\label{Fs_nf=3}
F_s(x,Q^2) &=& \frac{9}{2}F_{20}^{\gamma p}(x,Q^2)-\frac{1}{4}T_8(x,Q^2)-\frac{3}{8}U(x,Q^2), \quad n_f=3, \\
\label{Fs_nf=4}
F_s(x,Q^2) &=&  \frac{18}{5}F_{20}^{\gamma p}(x,Q^2) -\frac{1}{5}T_8(x,Q^2)+\frac{1}{5}T_{15}(x,Q^2)-\frac{3}{10}U(x,Q^2),  \quad n_f=4,  \\
\label{Fs_nf=5}
F_s(x,Q^2) &=& \frac{45}{11}F_{20}^{\gamma p}(x,Q^2) -\frac{5}{22}T_8(x,Q^2)+\frac{5}{22}T_{15}(x,Q^2)-\frac{3}{22}T_{24}(x,Q^2) -\frac{15}{44}U(x,Q^2), \quad n_f=5.
\ea
Here $F_{20}^{\gamma p}$ is an expression of LO form in terms of the quark distributions,
\be
\label{F20_defined}
F_{20}^{\gamma p} = \sum_{i=1}^{n_f} e_i^2x(q_i+\bar{q}_i),
\ee
with the sum running over the active quarks. We will  be concerned later with values of $Q^2$ above the $b$-quark excitation threshold at $m_b^2$, but not so large that $t$-quark effects are significant, so will generally take $n_f=5$ in the following discussion.

The measured structure function $F_2^{\gamma p}$  is related to $F_{20}^{\gamma p}$ by convolution with QCD corrections from the operator product expansion \cite{bardeen,hw,furmanski},
\be
\label{F2_F20_connection}
x^{-1}F_2^{\gamma p}= \left[\openone+\frac{\alpha_s}{2\pi}C_{2q}\right]\otimes \left(x^{-1}F_{2,0}^{\gamma p}\right) + \frac{\alpha_s}{2\pi}\Big(\sum_ie_i^2\Big)C_{2g}\otimes g,
\ee
where $\openone$ is the unit operator and the convolution  $\otimes$ is defined in \eq{convolution}. The coefficient functions $C_{2q}$ and $C_{2g}$ are given in \cite{esw} to NLO. Conversely,
\be
\label{F20}
x^{-1}F_{20}^{\gamma p} =  \left[\openone+\frac{\alpha_s}{2\pi}C_{2q}\right]^{-1}\otimes \left(x^{-1}F_2^{\gamma p} -  \frac{\alpha_s}{2\pi}\Big(\sum_ie_i^2\Big)C_{2g}\otimes g\right).
\ee

The inverse operator can be evaluated using Laplace transforms as discussed in Appendix \ref{Appendix:F20}.  We will assume that $F_{20}^{\gamma p}$ is known.

It is useful to note that $T_{15}$ and $T_{24}$ are given directly at  $Q^2=m_c^2,\, m_b^2$, the thresholds at which the $c$ and $b$ become active and below which their distributions vanish, by the $F_s$ distributions for $n_f=3$ and 4 at those thresholds,
\be
\label{Ts_from_Fs}
T_{15}(x,m_c^2)=F_s(x,m_c^2), \quad T_{24}(x,m_b^2)=F_s(x,m_b^2).
\ee
 In particular, the $x$ dependence of the $T_{15}$ and $T_{24}$\ is determined by $F_s$ at the thresholds.

We can use the expressions above to solve for the $s$, $c$, $b$, and light-quark distribution functions at small $x$ where we can ignore valence effects and the very small splittings between the $u$ and $d$ distributions generated by $V_1$, $V_2$, and $T_3$  and set those functions equal to zero. Then with $xq_\ell$ denoting the common small-$x$ distribution function $\bar{u}=\bar{d}$, and with $\bar{s}=s$, $\bar{c}=c$, and $\bar{b}=b$, the quark distributions for $n_f=5$ are
\ba
\label{s}
xs &=& xq_\ell-\frac{1}{4}T_8, \\
\label{c}
xc &=& xq_\ell-\frac{1}{12}T_8-\frac{1}{6}T_{15}, \\
\label{b}
xb &=& xq_\ell-\frac{1}{12}T_8-\frac{1}{24}T_{15}-\frac{1}{8}T_{24}; \\
\label{q_ell}
xq_\ell &=& \frac{1}{10}\left(F_s+\frac{5}{6}T_8+\frac{5}{12}T_{15}+\frac{1}{4}T_{24}-\frac{15}{4}U\right).
\ea
%

%%%%%%%%%%%%%%%%%%%%%%%%%%%%%%%%%%%%%%%%%
 %%%%%%%%%% SEC. QUARK DISTRIBUTIONS %%%%%%%%%%%%%%%%

% \section{ Quark distributions at very small $x$  \label{sec:quark_distributions}}

 %%%%%%%%%% SUBSEC. RESULTS FOR QUARKS %%%%%%%%%%%%%%%%%

 \subsection{Results \label{subsec:results_for_quarks}}

  %%%%%%%%%%%%%%%%%%%%%%%%%%%

We will use the expressions in Eqs.\ (\ref{s})-(\ref{q_ell})  to determine the behavior of the quark parton distribution functionns (PDFs) at ultra-low $x$ and large $Q^2$ implied by our  Froissart-bounded model for $F_2^{\gamma p}$.  This requires that we know $G(x,Q^2)=xg(x,Q^2)$ for use in \eq{F20}, as well as $U(x,Q^2)$ and $T_8(x,Q^2)$.
We will take $G$, $U$ and $T_8$ from existing parton-level  fits to the HERA data, with $G$ and $T_8$ extrapolated to small $x$ using quadratic expressions in $v=\ln(1/x)$,  a form implied by the  DGLAP evolution equations \cite{dglap1,dglap2,dglap3} for an input $F_s$ quadratic in $v$ \cite{bdm1,bdm2,bdhmapp}.

Given this input, we can determine $F_{20}^{\gamma p}(x,Q^2)$ from our fit to the HERA data using \eq{F20}, and $F_s(x,Q^2)$ for $n_f=3$, $Q^2\leq m_c^2$ from \eq{Fs_nf=3}. The latter, evaluated at $Q^2=m_c^2$, determines $T_{15}(x,m_c^2)$ through \eq{Ts_from_Fs}. We can then calculate the evolution of $T_{15}(x,Q^2)$ to $Q^2=m_b^2$  using the results in \cite{bdhmNLO}, and then repeat the process to obtain $T_{24}$. As we show in Appendix \ref{Appendix:NSevolution}, the effects of non-singlet QCD evolution are small so that $T_{15}(x,Q^2) \approx T_{15}(x,m_c^2)$ and $T_{24}(x,Q^2) \approx T_{24}(x,m_b^2)$. We also derive approximate analytic expressions for the evolved functions there; these are valid at large $v$.

We will only look at the quark distributions in the region of large $Q^2$, so will take $n_f=5$. Then from \eq{Fs_nf=5}, $F_s$ is given in $v$ space by
\be
\label{Fs_approx}
\hat{F}_s = \frac{45}{11}\hat{F}_{20}^{\gamma p}-\frac{5}{22}\hat{T}_8+\frac{5}{22}\hat{T}_{15}-\frac{3}{22}\hat{T}_{24}-\frac{15}{44}\hat{U}, \quad n_f=5,
\ee
where $\hat{T}_i$ and $\hat{U}$ are $T_i$ and $U$ evaluated in  $v$ space, with $x\rightarrow e^{-v}$.

We have determined $\hat{F}_{20}^{\gamma p}$ from $\hat{F}_2^{\gamma p}$ using the NLO transformation in \eq{F20} as described in the Appendix. The results are given analytically for large $v$ in \eq{F20example}. We used a gluon distribution $\hat{G}$ obtained from a  fit to the CT10 gluon distribution \cite{CT10,Durham} of the form in \eq{F2hat_v}, quadratic in $v$ and $\ln Q^2$.  The expression for $\hat{G}$ was fitted over the region $2\times 10^{-4}\leq x\leq0.01$ and 10 GeV$^2\leq Q^2\leq 1000$ GeV$^2$, and then extended to all $v$, $Q^2$.  We note that the CT10 gluon distributions obtained in NLO and NNLO are very similar, and agree also with the HERAPDF results \cite{HERAcombined,Durham}.

The resulting $\hat{F}_{20}^{\gamma p}$, with the transformation in \eq{F20} calculated in NLO, is compared with $\hat{F}_2^{\gamma p}$ in Fig.\ \ref{fig:F20F2comp}. The changes are on the order of 5-10\%, with a much smaller uncertainty from the gluon term. While we regard it as unlikely that higher order contributions to the functions $C_{2q}$ and $C_{2g}$ in \eq{F20} would affect the results for $\hat{F}_{20}^{\gamma p}$ significantly, we emphasize that any effects would be in the individual quark distributions, and would be insignificant for the relations between $\hat{F}_2^{\gamma p}$ and the corresponding structure functions $\hat{F}_2^{\nu(\bar{\nu})}$ and $\hat{F}0_2^{\nu(\bar{\nu})}$ for charged- and neutral-currents neutrino-nucleon scattering discussed in Sec.\ \ref{sec:wee_partons}.
%
%%%%%%%% FIG. 1 %%%%%%%%%%
%
\begin{figure}[htbp]
\includegraphics{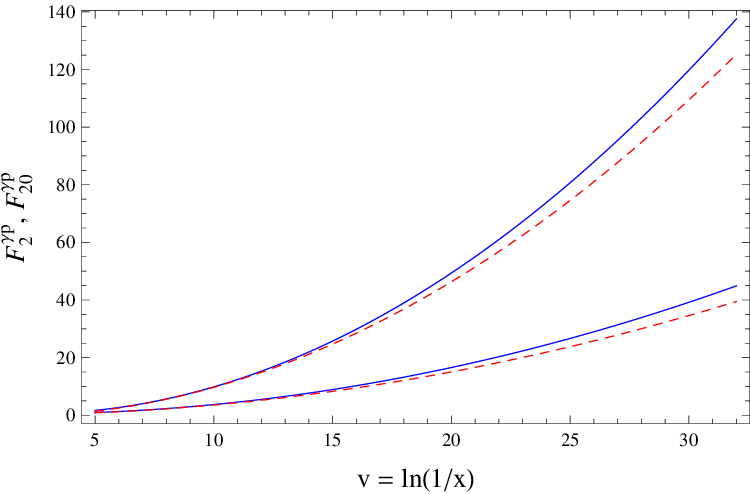}
\caption{Comparison of the quark-level distribution $\hat{F}_{20}^{\gamma p}$ in $v$ space (red dashed curves) with the large-$v$ extension of our Froissart form fit to the HERA data, \eq{F2hat_v} (solid blue curves), for $Q^2=10^4$ GeV$^2$ (top curves) and $Q^4=100$ GeV$^2$ (bottom). The two are related by \eq{F20} as implemented in the Appendix through the relation in \eq{F20example}. \label{fig:F20F2comp}}
\end{figure}
%%%%%%%%%%%%%%%%%%%%%
%

To get an approximate extension of the (small) function $\hat{T}_8$ to small $x$ or large $v$, we use a quadratic fit  to $\hat{T}_8$ as a function of $v$ as determined from the CT10 PDFs \cite{CT10,Durham} over the region
$10^{-5}<x<0.003$ for $Q^2=m_c^2$, and calculate it for larger $Q^2$ using the expression in \eq{FNSevolved}. We then determine $\hat{T}_{15}$ and $\hat{T}_{24}$ at the $c$ and $b$ thresholds $Q^2=m_c^2,\,m_b^2$ using the expressions in \eq{Ts_from_Fs}, again evolved to higher $Q^2$ using  \eq{FNSevolved}.
The relations in Eqs.\ (\ref{s}) to (\ref{q_ell}) then determine the $s$, $c$, $b$, and light quark distributions in terms of $\hat{F}_{20}^{\gamma p}$.

The results for the quark distributions are shown in Fig.\ \ref{Fig:QuarkPlot} for $5\leq v \leq 32.2$, corresponding to $0.0067 \geq x \geq 1\times 10^{-14}$. For comparison, the lower limit of the HERA data for $Q^2$ of a few GeV$^2$ is on the order of $x=10^{-4},\ v=9.2$.

%%%%%%% FIG. 2 %%%%%%%%%%%
%
\begin{figure}[htbp]
\includegraphics{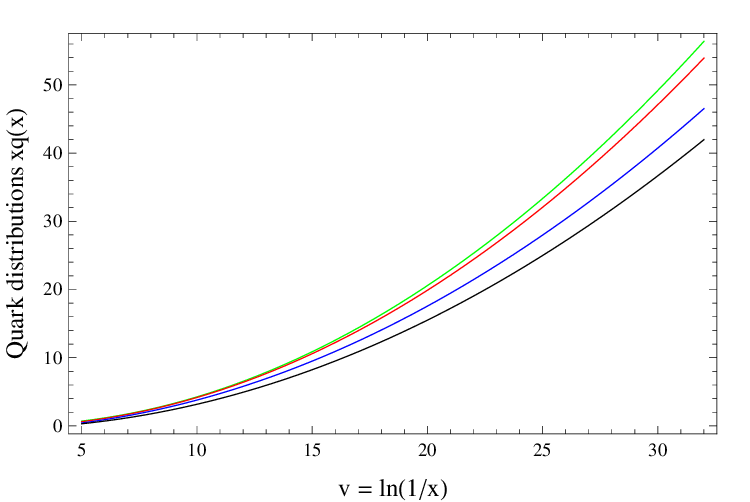}
\caption{Plots of quark distributions determined from the Froissart-bounded fit to $F_2^{\gamma p}$ versus $v=\ln(1/x)$ for $Q^2=10^4$ GeV$^2$:
 top to bottom, $xq_\ell(x,Q^2)$ (green curve), $xs(x,Q^2)$ (red curve), $xc(x,Q^2)$ (blue curve), and $xb(x,Q^2)$ (black curve). \label{Fig:QuarkPlot}}
\end{figure}
%%%%%%%%%%%%%%%%%%%%%
%

These distributions can be fitted to an accuracy  $\lesssim 1$\%, mostly much better, for $100\leq Q^2\leq10,000$ GeV$^2$, $5\leq v\leq 30$, by expressions of the same form as used for $F_2^{\gamma p}$ at small $x$,
\be
\label{quark_fit_func}
xq(x,Q^2) = \sum_{n,m=0}^2 A_{n,m}\ln^n(1/x)\ln^m{Q^2}.
\ee
  The errors in the $c$ distribution reach a few percent for $Q^2\lesssim 30$, a region that is completely unimportant in the  calculations in the next section.  The coefficients are given in Table\  \ref{tab:q_parametrization}.

%
%%%%%%% TABLE 2 %%%%%%%%%%%
%%%%%%%%%%%%%%%%%%%%%%%%%%
\begin{table}[htbp]
   \begin{center}
   \renewcommand{\arraystretch}{1.5}
   \begin{tabular}{|c|c|c|c|c|c|c|c|c|c|}
         \hline
 Quark& $A_{0,0}$ & $A_{0,1}$ & $A_{0,2}$ &$A_{1,0}$  & $A_{1,1}$ & $A_{1,2}$ & $A_{2,0}$ & $A_{2,1}$ & $A_{2,2}$\\ \hline\hline
 $q_\ell$ & \, 0.7616\, &\, -0.09057\, &  0.003667 & \,-0.1784\, &\,0.03601\, &\,-0.004328\, & \,0.1466\, &\,-0.001150\, & \,0.0006773\, \\ \hline
 $s$ &  0.51362 &  -0.09743 & 0.003897  &  -0.12414 & 0.03704 & -0.004366 & 0.01066 & -0.001155 & 0.0006776\\ \hline
 $c$ &  0.07983 &  -0.1107 & 0.004326  & -0.02167  & 0.03961 & -0.004456 & 0.0002307 & -0.001155 & 0.0006776\\ \hline
 $b$ & 0.2902  & -0.1092  &  0.0044717 & -0.07610  & 0.03833 &-0.004341  & -0.002686 & -0.001146 &  0.0006766 \\ \hline
  \end{tabular}
   \renewcommand{\arraystretch}{1}
   \end{center}
   \caption{Parameters in the fits to the quark distributions in \eq{quark_fit_func}. The distributions were fitted over the range 2 GeV$^2\leq Q^2\leq 50,000$ GeV$^2$. The $c$ distribution vanishes identically for $Q^2<M_c^2$, the $b$ distribution, for $Q^2<M_b^2$.}
   \label{tab:q_parametrization}
\end{table}
%%%%%%%%%%%%%%%%%%%%%%%%%%

 %%%%%%%% SEC. neutrino results %%%%%%%%%%%%%%%
 %%%%%%%%%%%%%%%%%%%%%%%%%%%%%%%%%%%%%

 \section{An application: Neutrino structure functions and the wee parton limit \label{sec:wee_partons}}

 %%%%%%%%%%%%%%%%%%%%%%%%%%%

The assumption that the differences between the quark distributions tend toward zero for small $x$ and large $Q^2$---Feynman's wee parton picture---was used in  \cite{bbmt,bhm} to calculate the neutrino and antineutrino cross sections on an isoscalar nucleon $N=(p+n)/2$ at very high energies in terms of $F_2^{\gamma p}$, neglecting QCD corrections.
The results derived  here show that there is {\em not} a proper wee parton limit, contrary to that assumption. This is clear  from Fig.\ \ref{Fig:QuarkPlot}.

As shown in Appendix\ \ref{Appendix:F20}, the quadratic large-$v$ behavior of our Froissart-type $F_2^{\gamma p}$ is preserved under the transformation to $F_{20}^{\gamma p}$ in \eq{F20}. The $T_i$ determined from $F_s$ at threshold values of $Q^2$ therefore share this behavior and must diverge as $v^2$ with increasing $v$. As a result, Eqs.\ (\ref{s})-(\ref{b}) show that the $s$, $c$ and $b$ distributions diverge from the light-quark distribution $q_\ell$ and from each other with increasing $v$, but approach constant ratios for $v$ large and $Q^2$ fixed as seen in Fig.\ \ref{Fig:QuarkRatios}, and the wee parton picture fails. The wee limit fails in general for cross sections $F_2^{\gamma p}$ which diverge as $x\rightarrow 0$ at fixed $Q^2$.
%
%%%%% FIG. 3 %%%%%%%%%%
%
\begin{figure}[htbp]
\includegraphics{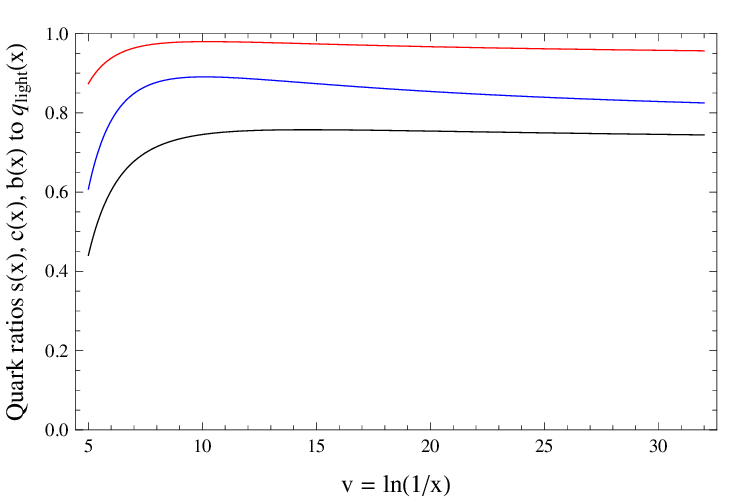}
\caption{The ratios, top to bottom, of the distributions $xs/xq_\ell$ (red curve), $xc/xq_\ell$ (blue curve), and $xb/xq_\ell$ (black curve) plotted versus $v=\ln(1/x)$ for $Q^2=10^4$ GeV$^2$. \label{Fig:QuarkRatios}}
\end{figure}
%%%%%%%%%%%%%%%%%%%%%%
%

Perhaps surprisingly, this result does not affect the supposed ``wee parton'' result for the neutrino cross sections in practice. The dominant structure function in charged-current neutrino scattering is $F_2^{\nu(\bar{\nu})}$ is given in terms of quark distributions for $n_f=5$ by
\ba
\label{F2nu1}
F_{20}^\nu &=& x(u+d+2s+2b+\bar{u}+\bar{d}+2\bar{c})  \\
\label{F2nu2}
&=& F_s\\
\label{F20nuF20relation}
&=& \frac{45}{11}F_{20}^{\gamma p} -\frac{1}{22}\left(5T_8-5T_{15}+3T_{24}\right)-\frac{15}{44}U.
\ea
The results for antineutrino scattering are identical, $F_2^{\bar{\nu}}=F_2^\nu$. We note that the coefficient 45/11=10/(22/9) of $F_{20}^{\gamma p}$ in this expression is just the ratio $\big(\sum_ic_{2,i}\big)\big/\big(\sum_i e_i^2\big)$ of the coefficients $c_{2,i}$ of the quark distributions  in \eq{F2nu1} to the sum of the squares of the quark charges; the same relation holds for the coefficients of $F_{20}^{\gamma p}$ in Eqs.\ (\ref{Fs_nf=3})-(\ref{Fs_nf=5}), with the sums in each case running over the active quarks.

We obtain the observable neutrino structure function $F_2^{\nu(\bar{\nu})}$ by applying the QCD corrections from the operator product expansion \cite{bardeen,hw,furmanski,esw} to $F_{20}^{\nu(\bar{\nu})}$
\be
\label{F2nu_F20nu_connection}
x^{-1}F_2^{\nu(\bar{\nu})}= \left[\openone+\frac{\alpha_s}{2\pi}C_{2q}\right]\otimes \left(x^{-1}F_{2,0}^{\nu(\bar{\nu})}\right) + \frac{\alpha_s}{2\pi}\Big(\sum_ic_{2,i}\Big)C_{2g}\otimes g,
\ee
where the  coefficient functions $C_{2q}$ and $C_{2g}$ are given to NLO in \eq{C_{2q}} and \eq{C_{2g}}.  This transformation is the inverse of that in \eq{F20} up to multiplication by  $\big(\sum_ic_{2,i}\big)\big/\big(\sum_i e_i^2\big)$, and converts the $F_{20}^{\gamma p}$ term in \eq{F20nuF20relation} to $F_2^{\gamma p}$ with the same coefficient. This relation holds to all orders in the strong coupling. It is a general QCD relation. The gluon term in \eq{F2nu_F20nu_connection} is absorbed in the process, and only $C_{2q}$ acts on the $T_j$ in \eq{F20nuF20relation}. The result in $v$ space for $Q^2>m_b^2$, $n_f=5$, is
\be
\label{F2nu_F2_connection_v}
\hat{F}_2^{\nu(\bar{\nu})} = (45/11)\hat{F}_2^{\gamma p}-\frac{1}{22}\left(5\hat{T}'_8-5\hat{T}'_{15}+3\hat{T}'_{24}\right)-\frac{15}{44}\hat{U}'.
\ee

The original functions    $\hat{T}_i(v,Q^2)$ in $v$ space are quadratic polynomials in $v$. The transformed functions  $\hat{T}'_i(v,Q^2)$ are again quadratics  for $v$ large. Their calculation is discussed in the Appendix\ \ref{Appendix:F20} around \eq{T_transformed1} where we show that, up to a small additive constant, $\hat{T}'(v,Q^2)$ is simply  $\hat{T}$ evaluated at a shifted value of $v$,
\be
\label{T'}
\hat{T}'(v,Q^2) = \hat{T}(v_T,Q^2) + {\rm constant} +O\left(e^{-v}\right), \quad v_T=v+{\rm constant}.
\ee

In the  wee parton limit $u=\bar{u}=d=\bar{d}=s =c=b=q_\ell$, the functions $T'_i$ and $U'$ in \eq{F2nu_F2_connection_v} vanish, and $F_{2,wee}^{\nu(\bar{\nu})}=(45/11)F_2^{\gamma p}$, a relation used in the case $n_f=4$ in the calculations of neutrino cross section in \cite{bbmt,bhm}. As seen in Fig.\ \ref{Fig:F2nuF2comp}, the results for $\hat{F}_2^{\nu(\bar{\nu})}$  obtained in this limit  for $n_f=5$ agree very well  for large $v$ with those  calculated using \eq{F2nu_F2_connection_v}}, for example, to $\sim 3.2\%$ (1.2\%) at $v=12$ (32) and $Q^2=100$ GeV$^2$, with the errors  decreasing with increasing $Q^2$ to 1.1\% (0.4\%) at $v=12$ (32)  for $Q^2=10,000$ GeV$^2$, even though the wee limit does not really exist for the quark distributions derived here. These differences are just discernible in Fig.\ \ref{Fig:F2nuF2comp}, and are not significant for applications at very small $x$ or large $v$.

It is not obvious that the relation  $F_{2}^{\nu(\bar{\nu})}\approx F_{2,wee}^{\nu(\bar{\nu})}=(45/11)F_2^{\gamma p}$ should hold as well as it does. In particular, the results in Fig.\ \ref{Fig:QuarkRatios}  show that the $c$ and $b$ PDFs are significantly smaller at all $v$ than the light-quark PDF $q_\ell$, while the valence distribution $U=u_v\approx 2d_v$ vanishes at large $v$. However, $F_2^{\gamma p}$ is fixed by experiment. The overall decrease in the contributions of the $s$, $c$ and $b$ quarks to   $F_2^{\gamma p}$ is therefore compensated by an increase in $q_\ell$. The  $c$ quark also appears with 4 times the weight of the  $s$ and $b$ quarks in $F_2^{\gamma p}$, but equal weight in $\hat{F}_2^{\nu(\bar{\nu})}$, with the result that the different errors in $s+b$ and $c$ tend to cancel in the latter; somewhat accidentally, the cancellation in nearly complete.

%
%%%%%% FIG. 4 %%%%%%%%%%%%%%%
%
\begin{figure}[htbp]
\includegraphics{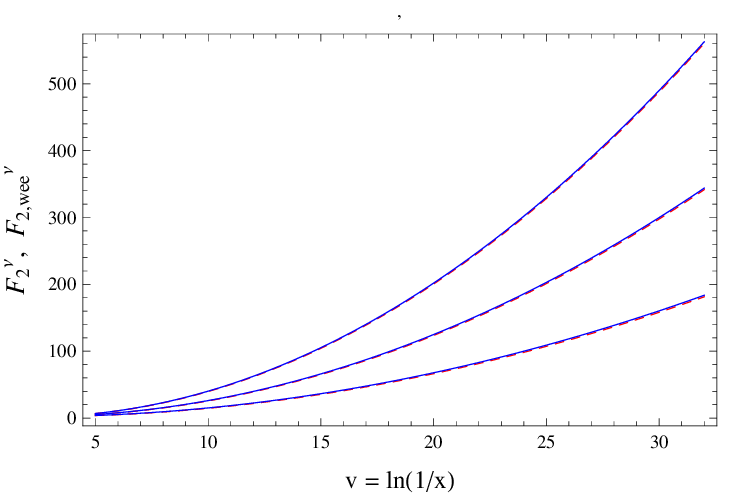}
\caption{Comparison of the dominant structure function $F_2^{\nu(\bar{\nu})}$ in charged current $\nu N$ or $\bar{\nu} N$ scattering for $n_f=5$ calculated for, top to bottom, $Q^2=10,000$, 1000, and 100 GeV$^2$ using the complete expression in \eq{F2nu_F2_connection_v} (dashed red curves), and the approximate distributions $F_{2,wee}^\nu\approx (45/11)F_2^{\gamma  p}$ (solid blue curves) derived assuming the validity of the wee parton limit for the quark distributions. The limits $v=5$ (32) of the range shown correspond to $x=0.007\  (10^{-14})$. \label{Fig:F2nuF2comp}}
\end{figure}
%%%%%%%%%%%%%%%%%%%%%%%%%%
%

\begin{figure}[htbp]
\includegraphics{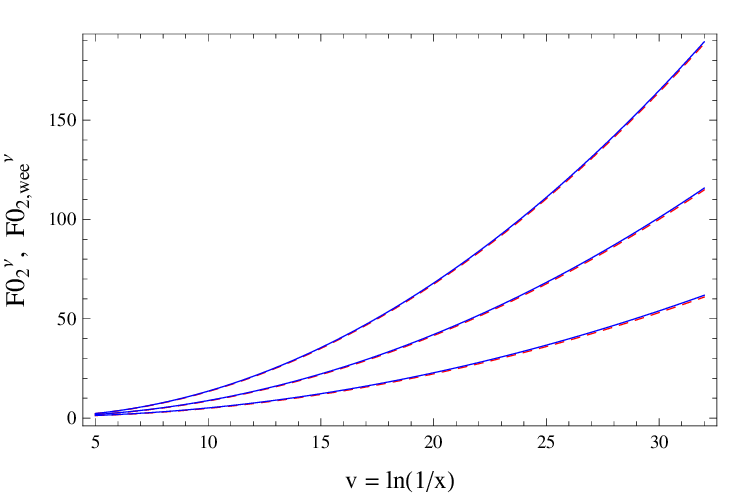}
\caption{Comparison of the dominant structure function $F0_2^{\nu(\bar{\nu})}$ in neutral current $\nu N$ or $\bar{\nu} N$ scattering calculated for, top to bottom, $Q^2=10,000$, 1000, and 100 GeV$^2$ using the complete expression (dashed red curves), and the approximate distributions $F0_{2,wee}^{\nu(\bar{\nu})}\approx (9/22)F_2^{\gamma  p}\left[3(L_d^2+R_d^2)+2(L_u^2+R_u^2)\right] $ (solid blue curves) derived assuming the validity of the wee parton limit for the quark distributions.  The limits $v=5$ (32) of the range shown correspond to $x=0.007\  (10^{-14})$. \label{Fig:F0nuF0nuweecomp}}
\end{figure}

Similar results hold for the structure function $F0_2^{\nu(\bar{\nu})}$  for neutral current $\nu N$ and $\bar{\nu} N$ scattering. The exact and (supposed) wee parton results  are compared Fig.\ \ref{Fig:F0nuF0nuweecomp}. The agreement is again very good for $v$ large, with agreement  to $\sim 3.8\%$ (1.5\%) at $v=12$, (32) for $Q^2=100$ GeV$^2$, decreasing with increasing $Q^2$ to 1.3\% (0.5\%) at $v=12$ (32)  for $Q^2=10,000$ GeV$^2$. These differences are quite small as seen in Fig.\ \ref{Fig:F0nuF0nuweecomp}, and are not significant for applications to ultra high energy neutrino cross sections.
%
%%%%%%% FIG. 5 %%%%%%%%%%
%
%%%%%%%%%%%%%%%%%%%%%%%%%%
%

{\em This agreement is important:}\/ the relation between $F_2^{\nu(\bar{\nu})}$  and the leading $F_2^{\gamma p}$ term in \eq{F2nu_F2_connection_v} holds to all orders in the strong coupling, and $F_2^{\gamma p}$ is determined directly by data. It can be extrapolated to large $v$ using a global Froissart-bounded fit to the $v$ and $Q^2$ dependence of the data {\em  independently of the individual quark distributions or the gluon distribution,}  eliminating uncertainties connected with the details of small-$x$ physics,  high-order QCD evolution, or the form of the initial parton distributions and how they are to be extrapolated. We emphasize, however, that relations derived utilizing the wee parton picture may, and do, fail in other situations. It is essential to check in each case.

Our full results for UHE neutrino cross sections are discussed in detail in the accompanying paper \cite{bdhmneutrino}.

%%%%%%%%%%%%%%%%%%%%%%%%%
%%%%%% SEC. CONCLUSIONS %%%%%%%%

\section{Conclusions \label{sec:conclusions}}

We have investigated the results that follow from the assumption that the cross section for the scattering of a virtual photon with $q^2=-Q^2<0$ from a nucleon, hence the structure function $F_2^{\gamma p}$ in deep inelastic $e^\pm p$ scattering, is hadronic in nature  with the same  Froissart-bounded structure as is observed in hadronic and real $\gamma p$ scattering. We have presented theoretical arguments in favor of this assumption, which is supported experimentally by the very accurate fit to the HERA data on $F_2^{\gamma p}$ obtained in earlier work. This fit is quadratic in the natural variable $v=\ln(1/x)$ for $x$ small, and allows a reliable extrapolation of $F_2^{\gamma p}$  to ultra-low values of $x$.

We have used this fit in conjunction with information on the small non-singlet function $T_8$ and the gluon distribution extrapolated consistently from results of the CT10 analysis of the HERA data \cite{CT10} to derive a complete set of quark distributions for $n_f=5$ active quarks for $x>10^{-14}$ (or $v<32$) and $x\lesssim 0.01$. The derivation does {\em not} use the DGLAP equations, which are expected to break down at very small $x$ \cite{GLRsmxqcd}. These quark distributions do not show the limiting behavior expected in the wee parton picture, in which the  deviations of the  distributions from one another tend to zero at small $x$ and large $Q^2$, but actually diverge from each other as $v^2=\ln^2(1/x)$ for $x\rightarrow 0$.

We show that, despite the failure of the wee parton picture at the quark level, the relations between $F_2^{\gamma p}$ and the dominant structure functions $F_2^{\nu(\bar{\nu})}$ and $F0_2^{\nu(\bar{\nu})}$ in charged- and neutral-current neutrino scattering derived in the wee parton picture continue to hold to  high accuracy at very small $x$. With this established, the use of the (supposed) wee parton relations to predict the dominant neutrino structure function $F_2^{\nu(\bar{\nu})}$ in terms of $F_2^{\gamma p}$ gives results that  hold to all orders in the strong coupling, and are {\em independent} of  assumptions about the  gluon distribution or the extrapolation of quark distributions characteristic of standard evolution-based methods.

The neutrino cross sections may be accessible at energies $E_\nu$ up to $10^{17}$ GeV in planned neutrino observatories, requiring values of $x$ down to $x=10^{-14}$. This corresponds to a relatively modest extrapolation by a factor of $\lesssim 3$ in $v$ from the upper values $v\sim 10$ explored at HERA to the values $v\sim 30$ needed for $E_\nu\sim 10^{17}$ GeV. We emphasize that, through the connections established here, measurements of the neutrino cross sections would allow, through the structure functions, the exploration of {\em hadronic} interactions at energies not otherwise accessible \cite{bdhmneutrino}.

As an important part of our analysis, we obtain simple analytic expressions in Appendices \ref{Appendix:NSevolution} and \ref{Appendix:F20} for the effects of non-singlet DGLAP evolution on the functions $T_8$, $T_{15}$, and $T_{24}$ needed in the derivation of quark distributions, and of the effects of the NLO QCD corrections needed to transform between the bare  $F_{20}^{\gamma p}$ expressed in terms of quark distributions, and the physical $F_2^{\gamma p}$. These are valid at large $v$ (small $x$) for structure functions with the Froissart-bounded form used here, and eliminate the need for the extensive numerical calculations commonly needed in $x$ space.

We conclude that the cross sections and quark distributions calculated from the small $x,$ large $Q^2$ extrapolation of $F_2^{\gamma p}(x,Q^2)$ from our saturated Froissart-bounded fit to the HERA data are the most physically motivated,  consistent with all other hadronic cross sections including $\gamma p$,  and provide the best estimate of the UHE energy neutrino-nucleon cross sections, which we develop fully in Part II \cite{bdhmneutrino}.

%%%%%%%%%%%%%%%%%%%%%%%

\begin{acknowledgments}

 M.\ M.\ B.\ and L.\ D.\ would like to thank the Aspen Center for Physics, where this work was supported in part by NSF Grant No. 1066293, for its hospitality.
 M.\ M.\ B.\  would like to thank Profs.\ A.\ Vainshtein and G.\ Domokos for valuable discussions.   P.\ H.\ would like to thank Towson University Fisher College of Science and Mathematics for support.  D.\ W.\ M.\ received support from DOE Grant No. DE-FG02-04ER41308.

\end{acknowledgments}

%%%%%%%%%%%%%%%%%%%%%%%%%%%

%%%%%%%%%%%%%%%%%%%%%%
%%%%%%%% APPENDIX %%%%%%%

\appendix

%%%%%%% APPENDIX A, ANALYTIC METHODS, NON-SINGLET EVOLUTION %%%%%%%%%%%

 \section{Non-singlet evolution and analytic methods \label{Appendix:NSevolution}}

 %%%%%%%%%%%%%%%%%%%%%%%%%%%

 In this Appendix, we develop analytic methods useful in treating the QCD evolution and transformation of the various functions considered in the main text. We take as our example the non-singlet quark distributions $T_i$; the more complicated transformations needed with $F_{20}^{\gamma p}$ and $F_2^{\nu(\bar{\nu})}$ are considered in Appendix \ref{Appendix:F20}. We emphasize that the methods used here give analytic results, valid for large $v$, for any distribution of the Froissart-bounded form in \eq{F2hat_v}. No numerical calculations are necessary, a great advantage at ultra-small $x$ relative to standard methods applied on a grid in $x$-$Q^2$ space.

 We find as an output  of the following analysis that the non-singlet $Q^2$ evolution of the functions $T_i$ is minimal. Those functions can therefore be approximated reasonably well at large $Q^2$ by their values at the threshold $Q_0^2=m_i^2$ where they are defined in terms of the singlet function $F_s$ or, up to small corrections, by the physical quantity $F_2^{\gamma p}$. We  also obtain simple analytic results which show that  the evolved $T_i$ can be obtained from the initial distributions by a small shift in the variable $v$ plus an additive constant.

 %%%%%%%%%%%%%%%%%%

 \subsection{Smallness of non-singlet evolution\label{subset:NSevolution}}

 %%%%%%%%%%%%%%%%%%

 Any of the non-singlet distributions evolves in $Q^2$ according to the equation (see Ref. \cite{bdhmNLO})
 \be
 \label{FNS_evolution}
 \hat{F}_{NS}(v,Q^2) = \int_0^v dw K_{NS}(w,Q^2)\hat{F}_{NS}(v-w,Q_0^2),
 \ee
 where $v=\ln(1/x)$, $\hat{F}_{NS}(v,Q^2)=F_{NS}(e^{-v},Q^2)$, and $K_{NS}$ is the evolution kernel
 \ba
 \label{KNS}
 K_{NS}(v,Q^2) &=& \frac{1}{2\pi i}\int_{-i\infty+\epsilon}^{i\infty+\epsilon}ds\, e^{vs}k_{NS}(s,Q^2), \\
 \label{kNS}
 k_{NS}(s,Q^2) &=& \exp\left[{\sum_n \tau_n \Phi_{NS}^{(n)}(s)}\right].
 \ea
 Here $\Phi_{NS}^{(n)}$ is the Laplace transform with respect to $v$ of the $n^{\rm th}$ order non-singlet splitting function in $x$, and
 \be
 \label{tau_n}
 \tau_n(Q^2,Q_0^2) = \left(\frac{1}{4\pi}\right)^n\int_{Q_0^2}^{Q^2} d(\ln\,Q'^2)\alpha_s^n(Q'^2).
 \ee

 In leading order ($n=1$), $\Phi_{NS}^{(1)}$ is just $\Phi_f$ as defined in Ref.  \cite{bdhmLO}, and can be written as
 \be
 \label{Phi_f}
 \Phi_{NS}^{(1)}(s) = \frac{4}{3}\left(\frac{2s}{s+1}+\frac{s}{s+2}\right)-\frac{16}{3}\left[\psi(s+1)-\psi(1)\right],
 \ee
 where $\psi(z)=\Gamma'(z)/\Gamma(z)$ is the digamma function.
 The function $ \Phi_{NS}^{(1)}(s) $ clearly vanishes at $s=0$. Its only singularities  in the complex $s$ plane are poles at $s=-1,\,-2,\ldots$, with the rightmost singularity at $s=-1$ (where here $s$ denotes a variable in Laplace space  and is {\em not} the Mandelstam invariant).  We can therefore move the contour of integration in \eq{KNS} to a line parallel to the imaginary axis with the real part of $s$ to the left of $s=0$, but still to the right of $s=-1$, without encountering any singularities.

The integrand diverges for $s\rightarrow -1$ and $s\rightarrow\infty$, and has its minimum value on the real axis  at the point $s_0\approx -1+\sqrt{8\tau_1/3v_1}$ near $-1$ where the derivative of the exponent with respect to $s$ vanishes. Here $v_1$ is given by $v_1\approx v-(8/9)(\pi^2-3)\tau_1$, so $v_1\rightarrow v$ for $v$ large. The integrand has a value proportional to $\exp[vs_0+\tau_1\Phi_{NS}^{(1)}(s_0)]\approx\exp[-v+\sqrt{32\tau_1v_1/3}+O(1/v_1)]$ at this saddle point. If we take the line of integration to run through the saddle point, we can estimate the integral using the method of steepest descents; the result is proportional to  $\exp[-v+\sqrt{32\tau_1v_1/3}]$.

 Given the behavior of the integrand,  we see that the integral is exponentially suppressed for $v\gg 32\tau_1/3$. As  a result, the kernel $K_{NS}^{(1)}$ is effectively zero except in a region  in $v$ extending only a distance $\sim 32\tau_1/3$ from $v=0$. This is small relative to $v$ for the values of $Q^2$ and $v$ of primary interest here. For example,  $\tau_1=0.108$  for $Q^2=10^4$ GeV$^2$, so the constraint requires only that $v\gg 1.2$ or $x\ll 0.3$. However, $v>11.5$ for $x<10^{-5}$, the region of primary interest, so the integral in \eq{FNS_evolution} samples only values of $\hat{F}_{NS}(v-w)$ very near $v$.

To exploit this observation,  we replace $w$ by zero in the factor $\hat{F}_{NS}(v-w,Q_0^2)$ in  \eq{FNS_evolution} and shift the contour in \eq{kNS} to the left of $s=0$. We find that
 \ba
 \hat{F}_{NS}(v,Q^2) &\approx& \hat{F}_{NS}(v,Q_0^2) \int_0^v dw K_{NS}(w,Q^2) \nonumber \\
   &=& \hat{F}_{NS}(v,Q_0^2) \int_0^v dw \,\frac{1}{2\pi i}\int_{-i\infty-c}^{i\infty-c}ds\, e^{ws}k_{NS}(s,Q^2) \nonumber \\
 \label{FNS_evolution_2}
&=& \hat{F}_{NS}(v,Q_0^2)\,\frac{1}{2\pi i}\int_{-i\infty-c}^{i\infty-c}ds\, \frac{1}{s}\left(e^{vs}-1\right)e^{\tau_1 \Phi_{NS}^{(1)}(s)}.
\ea
Since ${\rm Re}\, s<0$, the first term in the integrand is exponentially small for $v$ large and positive and can be dropped. We can close the contour to the right in the remaining term, and find that, for $v\gg 32\tau_1/3$ and $v\gg 1$,
\be
\label{FNSv_large}
\hat{F}_{NS}^{(1)}(v,Q^2) \approx \hat{F}_{NS}(v,Q_0^2)\,e^{\tau_1 \Phi_{NS}^{(1)}(0)} = \hat{F}_{NS}(v,Q_0^2)
\ee
where the last relation uses the fact that $ \Phi_{NS}^{(1)}(0)=0$. That is, {\em there is essentially no change in the NS distribution $\hat{F}_{NS}(v,Q_0^2)$ under LO evolution}.

This result  generalizes to higher orders: $\Phi_{NS}^{(n)}(s)$ has no singularities in $s$ to the right of $s=-1$ in NLO and NNLO, and presumably also higher orders, and $\Phi_{NS}^{(n)}(0)=0$ for all $n$, so, following the argument above, $\hat{F}_{NS}(v,Q^2)\approx\hat{F}_{NS}(v,Q_0^2)$ for $v$ sufficiently large.

The large-$v$ part of the argument is essentially unchanged.
To see that $\Phi_{NS}^{(n)}(0)=0$, we note that $\Phi_{NS}^{(n)}(s)$ is just the Laplace transform of the $n^{\rm th}$ order quark splitting function,
\be
\label{Phi_calc}
\Phi_{NS}^{(n)}(s) = \int_0^\infty dw\, e^{-w(s+1)}\hat{P}^-(w)=\int_0^\infty dw\, e^{-w(s+1)}\left(\hat{P}_{qq}(w)-\hat{P}_{q\bar{q}}(w)\right),
\ee
where we follow the notation in \cite{esw} with the substitution of $e^{-w}$ for $x$ and $\hat{P}^-(w)=P^-(e^{-w})$. For $s=0$, this reduces to
\ba
\label{Phi(0)}
\Phi_{NS}^{(n)}(0) &=& \int_0^\infty dw\, e^{-w}\left(\hat{P}_{qq}(w)-\hat{P}_{q\bar{q}}(w)\right) \nonumber \\
&=& \int_0^1dx\,\Big(P_{qq}(x)-P_{q\bar{q}}(x)\Big)=0.
\ea
This expression vanishes as the result of quark number conservation \cite{esw}. The insensitivity of any NS distribution $\hat{F}_{NS}(v,Q^2)$ to QCD evolution follows. We conclude that the result in \eq{FNSv_large} continues to hold, with negligible corrections because of the smallness of the higher- order parameters $\tau_n$ in \eq{kNS}, $\tau_n\ll\tau_1,\ n>1$.

%%%%%%%%%%%%%%

\subsection{Analytic methods \label{subsec:analytic_methods}}

%%%%%%%%%%%%%%

To see what residual effects of non-singlet evolution there are in a realistic case, we next treat the calculation above analytically. The same methods will be useful in Appendix\ \ref{Appendix:F20} in treating the more complicated cases encountered in the treatment of $F_{20}^{\gamma p}$ and the transformations of the $T_i$ under QCD corrections.

\begin{figure}[htbp]
\includegraphics{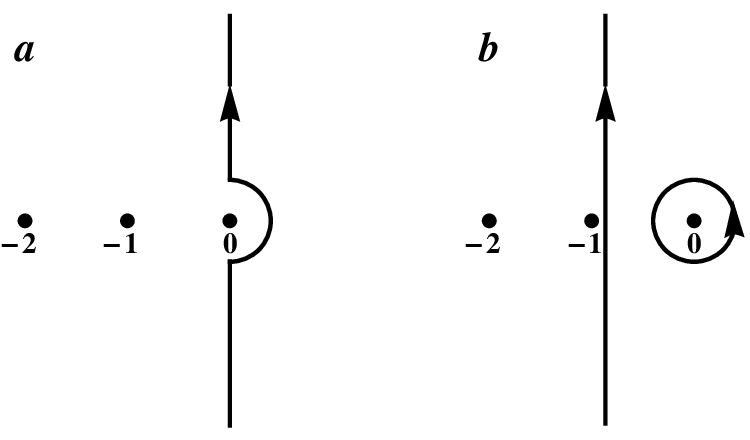}
\caption{Integration contours for the inverse Laplace transforms in \eq{FNSintegralasymp} and \eq{f_inverse}: (a), the original contour $(-i\infty+\epsilon,i\infty+\epsilon)$ which avoids the rightmost poles of the integrands at $s=0$ on the right; (b), the shifted contour, broken into a loop around the origin in the $s$ plane, and a line integral just to the right of the singularity at $s=-1$. That integral can be taken most efficiently to run through the saddle point near -1. There are further singularities at $s=-2,\dots,$ indicated by dots. \label{Fig:IntegrationContours}}
\end{figure}

We suppose that $\hat{F}_{NS}(v,Q_0^2)=\sum_{n=0}^2 c_n v^n$; this is the asymptotic form of  our Froissart-bounded fit to the HERA data, \eq{F2hat_v} at large $v$, and is also the form of  any of the non-singlet distributions listed above. It is useful in this case to use the alternative form of \eq{FNS_evolution} given by the convolution theorem for Laplace transforms,
\be
\label{FNS_Laplace_form}
\hat{F}_{NS}(v,Q^2) = {\cal L}^{-1}\left[k_{NS}(s,Q_0^2)f_{NS,0}(s);v\right],
\ee
where $f_{NS,0}(s)$ is the Laplace transform of the initial distribution $\hat{F}_{NS}(v,Q_0^2)$ at $Q_0^2$ with respect to $v$, $f_{NS,0}(s) =\sum_{n=0}^2c_n n!/s^{n+1}$.
In LO, this gives the evolved function
\be
\label{FNSintegralasymp}
\hat{F}_{NS}(v,Q^2)= \sum_{n=0}^2 c_n \frac{n!}{2\pi i}\int_{-i\infty+\epsilon}^{i\infty+\epsilon}\frac{ds}{s^{n+1}}\, e^{vs+\tau_1(Q^2,Q_0^2) \Phi_{NS}^{(1)}(s)}.
\ee
Since the only singularity of the integrand to the right of $s=-1$ is the pole  $1/s^{n+1} $ at $s=0$, we  can shift the integration contour to the left of $s=0$ as shown in Fig.\ \ref{Fig:IntegrationContours}, picking up the residue of the function $\exp[vs+\tau_1 \Phi_{NS}^{(1)}(s)]$ at the pole, and find that
\be
\label{v^n_evolved}
v^n\rightarrow \frac{d^n}{ds^n}e^{vs+\tau_1 \Phi_{NS}^{(1)}(s)}\Big |_{s=0} +  \frac{n!}{2\pi i}\int_{-i\infty-c}^{i\infty-c}\frac{ds}{s^{n+1}}\, e^{vs+\tau_1 \Phi_{NS}^{(1)}(s)}.
\ee
The line integral which remains can be taken to run through the slightly-shifted saddle point near $s=-1$, and  again gives a contribution which is exponentially small for $v$ large and can be dropped.
%
%%%%% FIG.6 %%%%%%%%%%%
%
%%%%%%%%%%%%%%%%%%
%

For an input distribution $v^2$, we get an evolved distribution
\ba
\label{v^2_evolved}
v^2 &\rightarrow& \left[v-\left(\frac{8\pi^2}{9}-\frac{10}{3}\right)\tau_1\right]^2 - \left(6+\frac{16}{3}\psi^{''}(1)\right)\tau_1 + O(e^{-v}) \\
\label{v^2_evolved2}
&=& \left(v-5.4397\tau_1\right)^2+6.8219\tau_1.
\ea
Similarly,
\be
\label{v_evolved}
v\rightarrow v-\left(\frac{8\pi^2}{9}-\frac{10}{3}\right)\tau_1 +O(e^{-v}) = v-5.4397\tau_1.
\ee
Finally, since $ \Phi_{NS}^{(1)}(0)=0$, a constant input function is unchanged in the evolution up to exponentially small terms.

Combining terms and expressing the result in terms of $\hat{F}_{NS}(v,Q_0^2)$, \eq{FNSintegralasymp} gives the evolved distribution
\ba
\label{FNSevolved}
\hat{F}_{NS}(v,Q^2) &=& \hat{F}_{NS}(v',Q_0^2) - c_2\left(6+\frac{16}{3}\psi^{''}(1)\right)\tau_1(Q^2,Q_0^2) + O\left(e^{-v}\right),  \\
\label{vprime}
 v'&=& v-\left(\frac{8\pi^2}{9}-\frac{10}{3}\right)\tau_1(Q^2,Q_0^2),
\ea
where $c_2\equiv \hat{C}_{2,NS}(Q_0^2)$ is the coefficient of $v^2$ in $\hat{F}_{NS}(v,Q_0^2)$.
The NS evolution has simply shifted $v$ by a small $\tau_1$-dependent constant and added a small constant term. For example, $v\rightarrow v'= v-0.587$ with an additive constant $0.736\,c_2$ for  $Q^2=10^4$ GeV$^2$, $Q_0^2=4.5$ GeV$^2$, $\tau_1=0.108$. The constant term can be neglected for the values of $v$ of primary interest here.    We find, therefore, that an excellent approximation for the complete evolved distribution is $\hat{F}_{NS}(v,Q^2) = \hat{F}_{NS}(v',Q_0^2)$.

%%%%%%%% APPENDIX B, F20, Ts %%%%%%%%%%%%

\section{Calculation of $F_{2,0}^{\gamma p}$ and the $T_i$ at large $v$ \label{Appendix:F20}}

%%%%%%%%%%%%%%%%%%%%%%%%%%%%%%%

We recall that the structure function $F_2^{\gamma p}$  is given in terms of the quark-level expression $F_{20}^{\gamma p}$ in \eq{F20_defined} by convolution with a set of coefficient functions from the operator product expansion  \cite{hw,furmanski},
\be
\label{F2NLO}
x^{-1}F_2^{\gamma p} = \left[\openone+ \frac{\alpha_s}{2\pi}C_{2q}\right]\otimes\left(x^{-1} F_{2,0}^{\gamma p}\right) +  \frac{\alpha_s}{2\pi}\Big(\sum_ie_i^2\Big)C_{2g} \otimes\,g,
\ee
where the convolution $\otimes$ of operators $A$ and $B$ is defined as
\be
\label{convolution}
A\otimes B = \int_x^1 \frac{dz}{z}A(x/z)B(z) = \int_x^1 \frac{dz}{z}A(z)B(x/z).
\ee
The operator $\openone$ in \eq{F2NLO} is the unit operator  and the sum over charges in the second term runs over active quarks and antiquarks.

The coefficient functions $C_{2q}$ and $C_{2g}$ depend on the renomalization scheme used in perturbative calculations and the order to which they are carried. We assume the use of the standard $\overline{\rm MS}$ scheme in which, at NLO \cite{esw},
\ba
C_{2q} &=& \frac{4}{3}\left[-\left(\frac{\pi^2}{3}+\frac{9}{2}\right)\delta(1-z)+2\left(\frac{\ln(1-z)}{1-z}\right)_+ - \frac{3}{2}\left(\frac{1}{1-z}\right)_+  \right. \nonumber \\
\label{C_{2q}}
&& \left.+3+2z - (1+z)\ln(1-z) -\frac{1+z^2}{1-z}\ln z\right], \\
\label{C_{2g}}
C_{2g} &=& \frac{1}{2}\left[\left((1-z)^2+z^2\right)\ln\frac{1-z}{z}-8z^2+8z-1\right].
\ea
The coupling $\alpha_s(Q^2)$ is to be evaluated at the same order.

The expression in Eq.\ (\ref{F2NLO}) is usually used to determine $x^{-1}F_2^{\gamma p}$ from the individual quark and gluon distributions found in fits to the DIS data. However, the relation can also be inverted to determine $x^{-1}F_{2,0}^{\gamma p}$ directly at a given order in $\alpha_s$ in terms of the observable structure function $x^{-1}F_2^{\gamma p}$ and a given gluon distribution $g(x,Q^2)$, i.e.,
\be
\label{F20_2}
x^{-1}F_{20}^{\gamma p} =  \left[\openone+\frac{\alpha_s}{2\pi}C_{2q}\right]^{-1}\otimes \left(x^{-1}F_2^{\gamma p} -  \frac{\alpha_s}{2\pi}\Big(\sum_ie_i^2\Big)C_{2g}\otimes g\right).
\ee
This is the result we need to obtain the singlet quark distribution $F_s$ and individual quark distributions as outlined in Sec.\ \ref{subsec:relations_for_distributions}.  As discussed there, $F_s$  is determined  (except at very low $Q^2$) by $F_{20}^{\gamma p}$ and the non-singlet functions $T_{15}$ and $T_{24}$, themselves related to $F_s$.

We sketch here the evaluation of the inverse operator and the final expression in \eq{F20_2} using Laplace transform. This requires several steps.  We first  multiply  by $x$ and use the second form of \eq{convolution} to recast \eq{F2NLO} in the form
\ba
\label{F2NLO2_1}
F_2^{\gamma p}(x,Q^2) &=& \left[\openone+\frac{\alpha_s}{2\pi}\left(zC_{2q}\right)\right]\otimes F_{20}^{\gamma p}+\frac{\alpha_s}{2\pi}+\Big(\sum_ie_i^2\Big)\left(zC_{2g}\right)\otimes G \\
\label{F2NLO2}
&=& F_{20}^{\gamma p}(x,Q^2) + \frac{\alpha_s}{2\pi}\int_x^1 \frac{dz}{z}\,\left[zC_{2q}(z)F_{20}^{\gamma p}(x/z,Q^2)+\Big(\sum_ie_i^2\Big)zC_{2g}(z)G(x/z,Q^2)\right],
\ea
where $G(x,Q^2)=xg(x,Q^2)$. We next transform the terms in \eq{C_{2q}} that involve distributions. Since $F(x/z)\equiv F_2^{\gamma p}(x/z,Q^2)\equiv 0$ for $z<x$, we may take the lower limit of integration in the convolution as 0, and write
\ba
\label{+defined1}
\int_x^1 dz\,F(x/z)\frac{1}{(1-z)_+} &=& \int_0^1 dz\,\frac{F(x/z)}{(1-z)_+} \\
\label{+defined2}
&\equiv& \int_0^1 dz\,\frac{F(x/z)-F(x)}{1-z} \\
\label{+defined3}
&=& F(x)\ln(1-x)+\int_x^1 dz\,\frac{F(x/z)-F(x)}{1-z} \\
\label{+defined4}
&=& F(x)\ln\frac{1-x}{x}+x\int_x^1 \left(\frac{F(y)}{y}-\frac{F(x)}{x}\right)\frac{dy}{y-x} \\
\label{+defined5}
&=& \int_0^v dw \ln\left(1-e^{-(v-w)}\right)\frac{\partial \hat{F}(w)}{\partial w},
\ea
where we have used the definition of the ``$+$'' operation in \eq{+defined2}, evaluated the integral on the interval $(0,x)$,  changed the integration variable $z$ to $y=x/z$ in \eq{+defined4}, and finally introduced the natural variables $v=\ln(1/x)$ and $w=\ln(1/y)$ and integrated by parts in \eq{+defined5} using a limiting procedure as sketched in  Ref.  \cite{bdhmLO}. The function $\hat{F}(w)$ is defined as $\hat{F}(w)\equiv \hat{F}_{20}^{\gamma p}(w,Q^2)=F_{20}^{\gamma p}(e^{-w},Q^2)$.

A similar calculation for the term in \eq{C_{2q}} proportional to $\left(\ln(1-z)/(1-z)\right)_+$ gives
\be
\label{+defined6}
\int_x^1 dz\,F(x/z)\left(\frac{\ln(1-z)}{(1-z)}\right)_+ = \int_0^v dw \ln^2\left(1-e^{-(v-w)}\right)\frac{\partial \hat{F}(w)}{\partial w}.
\ee

Using these results and transforming the remaining terms in \eq{F2NLO2} to $v$ space, we obtain the expression
\ba
\hat{F}_2^{\gamma p}(v,Q^2) &=& \hat{F}_{20}^{\gamma p}(v,Q^2) +\frac{\alpha_s(Q^2)}{2\pi}\left\{-\left(6+\frac{4}{9}\pi^2\right)\hat{F}_{20}^{\gamma p}(v,Q^2) + \int_0^v dw\, \hat{H}_q(v-w)\hat{F}_{20}^{\gamma p}(w,Q^2)  \right. \nonumber \\
&& \left. +\int_0^v dw \left[\frac{8}{3}\ln^2\left(1-e^{-(v-w)}\right) - 4\ln\left(1-e^{-(v-w)}\right)\right]\frac{\partial \hat{F}_{20}^{\gamma p}(w,Q^2)}{\partial w}\right\} \nonumber \\
\label{F2NLOv}
&& +\frac{\alpha_s(Q^2)}{2\pi}\left(\sum_ie_i^2\right)\int_0^v dw\,\hat{H}_g(v-w)\hat{G}(w,Q^2).
\ea
Here $\hat{G}(w,Q^2)=G(e^{-w},Q^2)$. The functions $\hat{H}_q$ and $\hat{H}_g$ are defined as
\ba
\label{hatHq}
\hat{H}_q(v) &=& e^{-v}C'_{2q}(e^{-v}), \\
\label{hatHg}
\hat{H}_g(v) &=& e^{-v}C_{2g}(e^{-v}),
\ea
where $C'_{2q}(z)$ contains the terms in $C_{2q}$ other than the delta function and the ``+''  terms treated above, i.e.,
\be
\label{Hq}
C'_{2q}(z)= \frac{4}{3}\left[3+2z-\frac{1+z^2}{1-z}\ln z-(1+z)\ln(1-z)\right].
\ee

The right-hand side of \eq{F2NLOv} is a sum of convolutions in $v$ space, and can be factored  by Laplace transformation into a sum of the products of the transforms of the functions in those convolutions,
\be
\label{F2Laplace}
f_2(s) = f_{20}(s)+\frac{\alpha_s}{2\pi}f_{20}(s)\left(-6-\frac{4}{9}\pi^2+h_{q1}(s)+s\,h_{q2}(s)\right)+\frac{\alpha_s}{2\pi}\Big(\sum_ie_i^2\Big)\tilde{g}(s)h_g(s).
\ee
Here $f_{20}(s)$, $f_2(s)$, and $\tilde{g}(s)$ are the Laplace transforms of $\hat{F}_{20}$, $\hat{F}_2$, and $\hat{G}$ with respect to $v$, with their $Q^2$ dependence suppressed,
\ba
\label{f20}
f_{20}(s) &=& {\cal L}\left[\hat{F}_{20}^{\gamma p}(v,Q^2);s\right], \\
\label{f2}
f_2(s) &=& {\cal L}\left[ \hat{F}_2^{\gamma p}(v,Q^2);s\right], \\
\label{g}
\tilde{g}(s) &=& {\cal L}\left[\hat{G}(v,Q^2);s\right],
\ea
while
\ba
h_{q1}(s) &=& {\cal L}\left[\hat{H}_q(v);s\right] \nonumber \\
\label{h_q1}
&=&\frac{4}{3}\left(\frac{H_{s+1}+3}{s+1} + \frac{H_{s+2}+2}{s+2} +\zeta(2,s+1) + \zeta(2,s+3)\right), \\
s\,h_{q2}(s) &=&s\, {\cal L}\left[ \frac{8}{3}\ln^2(1-e^{-v})-4\ln(1-e^{-v});s\right]  \nonumber \\
\label{h_q2}
&=& \frac{8}{3}\left(\frac{\pi^2}{6}   +\left(H_s\right)^2-\psi'(s+1)\right)+4H_s, \\
h_g(s) &=& {\cal L}\left[\hat{H}_g(v);s\right] \nonumber \\
 \label{h_g}
&=& -\frac{1}{2}\frac{H_s+1}{s+1}+\frac{H_{s+1}+4}{s+2}-\frac{H_{s+2}+4}{s+3}.
\ea
In these expressions, $H_s=\psi(s+1)-\psi(1)$, $\psi(s)=\Gamma'(s)/\Gamma(s)$, and $\zeta(2,s)=\sum_{k=0}^\infty(k+s)^{-2}$ is the Hurwitz generalized zeta function of degree 2. The factor $s$ which multiplies $h_{q2}(s)$ in \eq{F2Laplace} and \eq{h_q2} arises from the derivative of $\hat{F}_{20}^{\gamma p}$ in \eq{F2NLOv} and the relation ${\cal L}[\partial f(w)/\partial w;s]=s\,{\cal L}[f(w);s]$.

In the expression in \eq{F2Laplace}, $f_2(s)$ is known from our fit to the HERA data, and $g(s)$ is assumed also to be known, for example, from the extension of $G(x,Q^2)$ from earlier parton level fits to the data as extended to small $x$. Solving for $f_{20}(s)$, we find that
\be
\label{F20Laplace}
f_{20}(s) = \left[f_2(s)-\frac{\alpha_s}{2\pi}\Big(\sum_ie_i^2\Big)h_g(s)\tilde{g}(s)\right]\Big/\big[1+(\alpha_s/2\pi)d(s)\big],
\ee
where
\be
\label{d(s)}
d(s) = -6-\frac{4}{9}\pi^2+h_{q1}(s)+s\,h_{q2}(s).
\ee
Thus, inverting the Laplace transform in \eq{f20}, we find that
\ba
\hat{F}_{20}^{\gamma p} &=&{\cal L}^{-1}\left[f_{20}(s); v\right]  \nonumber \\
\label{F20solution}
&=& {\cal L}^{-1}\left[\frac{f_2(s)}{1+(\alpha_s/2\pi)d(s)} -\frac{\alpha_s}{2\pi}\Big(\sum_ie_i^2\Big)\frac{h_g(s)\tilde{g}(s)}{1+(\alpha_s/2\pi)d(s)};v\right].
\ea

The inverse Laplace transform in \eq{F20solution}  can be calculated simply analytically for $v$ large or $x$ small. In particular, in our Froissart bounded model, $\hat{F}_2^{\gamma p}(v,Q^2)$ and $\hat{G}(v,Q^2)$ are essentially quadratic polynomials in $v$ for $v>>1$ as in \eq{F2hat_v}. In the polynomial terms, $v^n\rightarrow n!/s^{n+1}$ under Laplace transformation. The exponentially small terms omitted in \eq{F2hat_v} give extra poles for $s\rightarrow -1,\,-2,\ldots$, and, if retained, lead only to terms of order $e^{-v}$ or smaller in the final result. The main contributions for $n_f=5$ and $v>>1$ are therefore given by integrals of the form
\be
\label{f_inverse}
\frac{n!}{2\pi i}\int_{-i\infty+\epsilon}^{i\infty+\epsilon}\frac{ds}{s^{n+1}}e^{vs}\frac{\left[1,\  (\alpha_s/2\pi)(22/9)h_g(s)\right]}{1+(\alpha_s/2\pi)d(s)}, \quad n=0,1,2,
\ee
where the numerator function in \eq{f_inverse} is 1 for the $f_2$ term in \eq{F20solution} and $(\alpha_s/2\pi)(22/9)h_g(s)$ for the $g$ term.

The numerator functions have no singularities in the complex plane to the right of $s=-1$.  The function $d(s)$ has second-order poles for $s\rightarrow -1,\,-2,\,-3,\ldots$, but these cause no problems. However, the complete  denominator function has a pair of complex conjugate zeros near and to the right of $-1$ where
\ba
(s+1)^2\left[1+(\alpha_s/2\pi)d(s)\right] &\rightarrow&  (s+1)^2\left[1-\frac{\alpha_s}{2\pi}\left(\frac{10}{3}+\frac{2\pi^2}{3}\right)\right]+\frac{4}{3}\frac{\alpha_s}{2\pi} \nonumber \\
\label{pole_equation}
&& +\frac{2}{9}\frac{\alpha_s}{2\pi}\left[4\pi^2-12-21\,\psi''(1)-24\,\zeta(3)\right](s+1)^3+\cdots,
\ea
leading (in order $\alpha_s$) to complex conjugate poles at
 \be
 \label{poles}
 s\approx -1+O(\alpha_s^2)\pm i\left[\frac{\alpha_s}{2\pi}\left(\frac{10}{3}+\frac{2\pi^2}{3}\right)\right]^{1/2}\left[(1+O(\alpha_s)\right],
 \ee
 e.g., at $s=-0.9962\pm0.1750\,i$ for $\alpha_s\left(M_Z^2\right)=0.118$. There are further pairs of conjugate poles near $s=-2,\,-3,\ldots$. These pole positions will be shifted slightly and new poles introduced when the coefficient functions $C_{2q}$ and $C_{2g}$ are evaluated to higher orders in $\alpha_s$, introducing higher order contributions in $1/(s+1)$, but the rightmost singularities from the (generalized) factor $1/\left[1+(\alpha_s/2\pi)d(s)\right]$ should remain very close to $s=-1$.

 We conclude that the contours of integration in \eq{f_inverse} can be shifted to the left in the complex $s$ plane as in Fig.\ \ref{Fig:IntegrationContours} to run through saddle points close to $s=-1$, but just to the right of the complex conjugate poles in \eq{poles}, picking up the residues of the integrands at $s=0$ and leaving a residual integral which is suppressed by a factor $\approx e^{-v}$,  very small for $v$ large. We drop the latter.

 The calculation of the residues of the poles at $s=0$ is straightforward. Thus, for the ``1'' term in \eq{f_inverse}, the quadratic form of our input function $\hat{F}_2^{\gamma p}(v)$ in $v$ is reproduced in $\hat{F}_{20}^{\gamma p}(v)$ with small shifts in $v$ and an added constant in the $v^2$ term as given below in Eqs.\ (\ref{F20example}) and (\ref{vf_vg}). The analytic forms of the coefficients in these expressions are known, but are too complicated to record here.  Similar results hold for the ``$g$'' term.

 Combining the results, we find that, for general values of $b(Q^2)\equiv
\alpha_s(Q^2)/2\pi$,
 \ba
 \hat{F}_{20}^{\gamma p}(v,Q^2) &=&\hat{F}_2^{\gamma p}(v_f,Q^2)  -(5.523 -17.665\, b)\, b\,\hat{C}_{2f}(Q^2) \nonumber \\
 \label{F20example}
&&- \frac{11}{27}\,b\,\hat{G} (v_g,Q^2)  +
 \frac{11}{27}\,b\,(7.549+5.523\,b-17.665\,b^2)\,\hat{C}_{2g}(Q^2)+O\left(e^{-v}\right),
 \ea
to NLO, where $\hat{C}_{2f}(Q^2)$ is the coefficient of $v^2$ in $F_2^{\gamma p}(v,Q^2)$,  Eqs.\ (\ref{F2hat_v}) and (\ref{C2}), and $\hat{C}_{2g}(Q^2)$ is the coefficient of the corresponding term in $\hat{G}(v,Q^2)$. The shifted arguments $v_f$ and $v_g$ are
\be
\label{vf_vg}
v_f = v -4.203\,b,\quad v_g = v - 4.623-4.203\,b.
\ee

 The main uncertainty in the overall result for $\hat{F}_{20}^{\gamma p}$ arises from the uncertainty in the gluon distribution. This was treated  using an extrapolation of the CT10 \cite{CT10} $G(x,Q^2)$ quadratic in $v$, with coefficients quadratic in $\ln Q^2$, fitted to the NNLO $G$ over the region $2\times 10^{-4}\leq x\leq0.01$, 10 GeV$^2\leq Q^2\leq 1000$ GeV$^2$. As noted earlier, this agrees very well with the HERAPDF version of $G$.

 We emphasize that the form of these results, with quadratics in $v$ transformed to quadratics up to exponentially small corrections, is quite general, the result simply of the calculation of residues at $s=0$, with all other singularities of the integrands, from either the kernel functions in Laplace space or the forms of $\hat{F}_2^{\gamma p}$ or $\hat{G}$ for $v\sim 0$, displaced at least to the vicinity of $s=-1$.

The calculation of the neutrino structure functions $F_2^{\nu(\bar{\nu})}$ and $F0_2^{\nu(\bar{\nu})}$ also requires the evaluation of the action of $\left[\openone+(\alpha_s/2\pi)C_{2q}\right]$ on the functions $T_8$, $T_{15}$, and $T_{24}$. In $v$ space, these are quadratics in $v$ for $v$ large. The resulting transformation of the powers $v^n$, $n=0,\,1,\,2$ is just the inverse of that associated with the ``1'' term in the  transformation $\hat{F}_2^{\gamma p}\rightarrow \hat{F}_{20}^{\gamma p}$ discussed above; $\hat{G}$ does not enter. Thus,
\be
 \label{T_transformed1}
 \hat{T}_i(v,Q^2)\rightarrow\hat{T}_i(v_T,Q^2)+(5.523 -17.665\, b)\, b\,\hat{C}_{2,T_i}(Q^2),\quad v_T=v+ 4.203\,b,
 \ee
 with $\hat{C}_{2,T_i}(Q^2)$ the coefficient of the quadratic term in $v$ in $T_i(v,Q^2)$. The functions $\hat{T}_i(v,Q^2)$ are given in terms of the initial distributions $T_i(v,Q_0^2)$ determined at $Q_0^2=m_c^2,\,m_b^2$ by the expression in Eqs.\  (\ref{tau_n}), (\ref{FNSevolved}), and (\ref{vprime}).

 The calculation of the complete neutrino cross sections also requires the structure functions $xF_3$ and $F_L$. These are given to NLO,  using the form analogous to that for $F_2^{\gamma p}$ in \eq{F2NLO2_1}, by
\ba
 \label{NLOxF3_2}
xF_3^{\nu(\bar{\nu})}&=& xF_{3,0}^{\nu(\bar{\nu})}+\frac{\alpha_s}{ 2 \pi}\left(zC_{3q}\right)\otimes \left(zF_{3,0}^{\nu(\bar{\nu})}\right), \\
 \label{FLNLO_2}
F_{L}^{\nu(\bar{\nu})}(x,Q^2) &=& \frac{\alpha_s}{2\pi}\left(z C_{Lq}\right)\otimes F_{20}^{\nu(\bar{\nu})}+\frac{\alpha_s}{2\pi}2n_f\left(zC_{Lg}\right)\otimes G.
\ea
where, for example, $F_{3,0}^\nu=u+d+2s+2b-\bar{u}-\bar{d}-2\bar{c}$ for $n_f=5$. The coefficient functions are
\ba
\label{C3qC3g}
C_{3q}(z) &=& C_{2q}(z)-\frac{4}{ 3}(1+z) \label{C3q}, \quad C_{3g}=0\\
\label{CLqCLg}
C_{Lq}(z) &=& \frac{8}{3}z,\qquad C_{Lg}(z)=2z(1-z).
\ea

Transforming \eq{NLOxF3_2} to $v$ space and factoring the resulting convolution with a Laplace transform, we find that
\be
\label{F3Laplace}
f_3 = \left[\openone+b d(s) +b h_{3q}(s)\right]f_{30}.
\ee
where $f_3$ and $f_{30}$ are the Laplace transforms of $\hat{F}_3$ and $\hat{F}_{30}$ with respect to $v$, and
\be
\label{h3q}
h_{3q}(s)= -\frac{4}{3}\left(\frac{1}{s+1}+\frac{1}{s+2}\right).
\ee
Since $\hat{F}_{30}$ is a quadratic in $v$ for $v$ large, we can calculate the inverse Laplace transform of $f_3$ as above by  calculating the residues of the integrand $e^{vs} \left[\openone+b d(s) +b h_{3q}(s)\right] \left(n!/s^{n+1}\right)$ for $n=0,\,1,\,2$, corresponding to inputs $v^n$. The results give
\ba
\label{F3_F30_relation}
\hat{F}_3(v,Q^2) &=& (1-2 b)\hat{F}_{30}(v_3,Q^2)+\frac{(2.523-39.499 b)b}{1-2 b}\hat{C}_{2,3}(Q^2)+O\left(e^{-v}\right), \\
\label{v3}
v_3 &=& v+\frac{5.870 b}{1-2 b},
\ea
to NLO, with $\hat{C}_{2,3}(Q^2)$  the coefficient of $v^2$ in $\hat{F}_{30}(v,Q^2)$.

Similarly, for $F_L$, we find that
\ba
\label{f_L}
f_L &=& b h_{Lq}(s)f_{20}+2n_fbh_{Lg}(s)\tilde{g}, \\
\label{hLq_hLg}
h_{Lq} &=& \frac{8}{3}\frac{1}{s+2},\quad h_{Lg}=\frac{2}{s+2}-\frac{2}{s+3},
\ea
where $\tilde{g}$ is the Laplace transform of $\hat{G}(v,Q^2)$. Using the quadratic forms of $f_{20}$ and $\tilde{g}$ in $v$ and evaluating the residues at $s=0$ in the inverse Laplace transform, we get
\ba
\hat{F}_L(v,Q^2) &=& \frac{4\, b}{3}\hat{F}_{20}\left(v-\frac{1}{2},Q^2\right) + \frac{b}{3}\hat{C}_{2,f0} \nonumber \\
\label{FL_F2_G_relation}
&& + \frac{2n_fb}{3}\hat{G}\left(v-\frac{5 }{6},Q^2\right)+\frac{13n_f b}{54}\hat{C}_{2g}+O\left(e^{-v}\right),
\ea
in NLO, with $\hat{C}_{2,f0}(Q^2)$ and $\hat{C}_{2g}(Q^2)$  the coefficients of $v^2$ in $\hat{F}_{20}(v,Q^2)$ and $\hat{G}(v,Q^2)$.

%%%%%%%%%%%%%%%%%%%%%%0.0

\bibliography{small_x_references}

\begin{thebibliography}{48}
\expandafter\ifx\csname natexlab\endcsname\relax\def\natexlab#1{#1}\fi
\expandafter\ifx\csname bibnamefont\endcsname\relax
  \def\bibnamefont#1{#1}\fi
\expandafter\ifx\csname bibfnamefont\endcsname\relax
  \def\bibfnamefont#1{#1}\fi
\expandafter\ifx\csname citenamefont\endcsname\relax
  \def\citenamefont#1{#1}\fi
\expandafter\ifx\csname url\endcsname\relax
  \def\url#1{\texttt{#1}}\fi
\expandafter\ifx\csname urlprefix\endcsname\relax\def\urlprefix{URL }\fi
\providecommand{\bibinfo}[2]{#2}
\providecommand{\eprint}[2][]{\url{#2}}

\bibitem[{\citenamefont{Block et~al.}(2006)\citenamefont{Block, Berger, and
  Tan}}]{bbt1}
\bibinfo{author}{\bibfnamefont{M.~M.} \bibnamefont{Block}},
  \bibinfo{author}{\bibfnamefont{E.~L.} \bibnamefont{Berger}},
  \bibnamefont{and} \bibinfo{author}{\bibfnamefont{C.-I.} \bibnamefont{Tan}},
  \bibinfo{journal}{Phys. Rev. Lett.} \textbf{\bibinfo{volume}{97}},
  \bibinfo{pages}{252003} (\bibinfo{year}{2006}), \eprint{hep-ph/0610296}.

\bibitem[{\citenamefont{Berger et~al.}(2007)\citenamefont{Berger, Block, and
  Tan}}]{bbt2}
\bibinfo{author}{\bibfnamefont{E.~L.} \bibnamefont{Berger}},
  \bibinfo{author}{\bibfnamefont{M.~M.} \bibnamefont{Block}}, \bibnamefont{and}
  \bibinfo{author}{\bibfnamefont{C.-I.} \bibnamefont{Tan}},
  \bibinfo{journal}{Phys. Rev. Lett.} \textbf{\bibinfo{volume}{98}},
  \bibinfo{pages}{242001} (\bibinfo{year}{2007}), \eprint{hep-ph/0703003}.

\bibitem[{\citenamefont{Block et~al.}(2008P)\citenamefont{Block, Berger, McKay,
  and Tan}}]{bbmt}
\bibinfo{author}{\bibfnamefont{M.~M.} \bibnamefont{Block}},
  \bibinfo{author}{\bibfnamefont{E.~L.} \bibnamefont{Berger}},
  \bibinfo{author}{\bibfnamefont{D.~W.} \bibnamefont{McKay}}, \bibnamefont{and}
  \bibinfo{author}{\bibfnamefont{C.-I.} \bibnamefont{Tan}},
  \bibinfo{journal}{Phys. Rev. D} \textbf{\bibinfo{volume}{77}},
  \bibinfo{pages}{053007} (\bibinfo{year}{2008P}), \eprint{arXiv: 0708.1960v1
  [hep-ph]}.

\bibitem[{\citenamefont{Block et~al.}(2010{\natexlab{a}})\citenamefont{Block,
  Ha, and McKay}}]{bhm}
\bibinfo{author}{\bibfnamefont{M.}~\bibnamefont{Block}},
  \bibinfo{author}{\bibfnamefont{P.}~\bibnamefont{Ha}}, \bibnamefont{and}
  \bibinfo{author}{\bibfnamefont{D.}~\bibnamefont{McKay}},
  \bibinfo{journal}{Phys. Rev. D} \textbf{\bibinfo{volume}{82}},
  \bibinfo{pages}{077302} (\bibinfo{year}{2010}{\natexlab{a}}).

\bibitem[{\citenamefont{Block}(2006{\natexlab{a}})}]{blockrev}
\bibinfo{author}{\bibfnamefont{M.~M.} \bibnamefont{Block}},
  \bibinfo{journal}{Phys. Rep.} \textbf{\bibinfo{volume}{36}},
  \bibinfo{pages}{71} (\bibinfo{year}{2006}{\natexlab{a}}).

\bibitem[{\citenamefont{{ATLAS Collaboration}}(2011)}]{LHCtot1}
\bibinfo{author}{\bibnamefont{{ATLAS Collaboration}}}, \bibinfo{journal}{Nature
  Comm.} \textbf{\bibinfo{volume}{2}}, \bibinfo{pages}{463}
  (\bibinfo{year}{2011}).

\bibitem[{\citenamefont{Antchev et~al.}(2011)}]{LHCtot2}
\bibinfo{author}{\bibfnamefont{G.}~\bibnamefont{Antchev}} \bibnamefont{et~al.}
  (\bibinfo{collaboration}{TOTEM Collaboration}), \bibinfo{journal}{Euro. Phys.
  Lett.} \textbf{\bibinfo{volume}{96}}, \bibinfo{pages}{21002}
  (\bibinfo{year}{2011}).

\bibitem[{LHC()}]{LHCtot3}
\bibinfo{note}{CMS Collaboration, \uppercase{C}ERN Document Server,
  http://cdsweb.cern.ch/record/1372466?ln=en, 2011}.

\bibitem[{\citenamefont{Abreu et~al.}(2012)}]{POAp-air}
\bibinfo{author}{\bibfnamefont{P.}~\bibnamefont{Abreu}} \bibnamefont{et~al.}
  (\bibinfo{collaboration}{Pierre Auger Collaboration}),
  \bibinfo{journal}{Phys. Rev. Lett. 062002 (2012)}
  \textbf{\bibinfo{volume}{109}}, \bibinfo{pages}{062002}
  (\bibinfo{year}{2012}), \eprint{arXiv:1208.1520 [hep-ex]}.

\bibitem[{\citenamefont{Block}(2011)}]{mbair}
\bibinfo{author}{\bibfnamefont{M.}~\bibnamefont{Block}},
  \bibinfo{journal}{Phys. Rev. D} \textbf{\bibinfo{volume}{84}},
  \bibinfo{pages}{091501} (\bibinfo{year}{2011}).

\bibitem[{\citenamefont{Block and Halzen}(2011)}]{blockhalzen}
\bibinfo{author}{\bibfnamefont{M.~M.} \bibnamefont{Block}} \bibnamefont{and}
  \bibinfo{author}{\bibfnamefont{F.}~\bibnamefont{Halzen}},
  \bibinfo{journal}{Phys. Rev. Lett.} \textbf{\bibinfo{volume}{107}},
  \bibinfo{pages}{212002} (\bibinfo{year}{2011}).

\bibitem[{\citenamefont{Block and Halzen}(2012)}]{blockhalzen2}
\bibinfo{author}{\bibfnamefont{M.~M.} \bibnamefont{Block}} \bibnamefont{and}
  \bibinfo{author}{\bibfnamefont{F.}~\bibnamefont{Halzen}},
  \bibinfo{journal}{Phys. Rev. D} \textbf{\bibinfo{volume}{86}},
  \bibinfo{pages}{051504} (\bibinfo{year}{2012}).

\bibitem[{\citenamefont{{Abolleir Fernandez} et~al.}(2012)}]{LHeC}
\bibinfo{author}{\bibfnamefont{J.~L.} \bibnamefont{{Abolleir Fernandez}}}
  \bibnamefont{et~al.}, \bibinfo{journal}{J.\ Phys.\ G}
  \textbf{\bibinfo{volume}{39}}, \bibinfo{pages}{075001}
  (\bibinfo{year}{2012}).

\bibitem[{\citenamefont{Block et~al.}(2012)\citenamefont{Block, Durand, Ha, and
  McKay}}]{bdhmneutrino}
\bibinfo{author}{\bibfnamefont{M.~M.} \bibnamefont{Block}},
  \bibinfo{author}{\bibfnamefont{L.}~\bibnamefont{Durand}},
  \bibinfo{author}{\bibfnamefont{P.}~\bibnamefont{Ha}}, \bibnamefont{and}
  \bibinfo{author}{\bibfnamefont{D.~W.} \bibnamefont{McKay}}
  (\bibinfo{year}{2012}), \bibinfo{note}{the companion paper, this journal}.

\bibitem[{\citenamefont{Letinien et~al.}(2004)\citenamefont{Letinien, Gorham,
  Jacobson, and Roussel-Dupr\'{e}}}]{forte}
\bibinfo{author}{\bibfnamefont{N.}~\bibnamefont{Letinien}},
  \bibinfo{author}{\bibfnamefont{P.}~\bibnamefont{Gorham}},
  \bibinfo{author}{\bibfnamefont{A.}~\bibnamefont{Jacobson}}, \bibnamefont{and}
  \bibinfo{author}{\bibfnamefont{R.}~\bibnamefont{Roussel-Dupr\'{e}}},
  \bibinfo{journal}{Phys. Rev. D} \textbf{\bibinfo{volume}{69}},
  \bibinfo{pages}{013008} (\bibinfo{year}{2004}).

\bibitem[{\citenamefont{Gorham et~al.}(2004)}]{glue}
\bibinfo{author}{\bibfnamefont{P.}~\bibnamefont{Gorham}} \bibnamefont{et~al.}
  (\bibinfo{collaboration}{GLUE Collaboration}), \bibinfo{journal}{Phys. Rev.
  Lett.} \textbf{\bibinfo{volume}{93}}, \bibinfo{pages}{041101}
  (\bibinfo{year}{2004}).

\bibitem[{\citenamefont{Block et~al.}(2011{\natexlab{a}})\citenamefont{Block,
  Durand, Ha, and McKay}}]{bdhmapp}
\bibinfo{author}{\bibfnamefont{M.~M.} \bibnamefont{Block}},
  \bibinfo{author}{\bibfnamefont{L.}~\bibnamefont{Durand}},
  \bibinfo{author}{\bibfnamefont{P.}~\bibnamefont{Ha}}, \bibnamefont{and}
  \bibinfo{author}{\bibfnamefont{D.~W.} \bibnamefont{McKay}},
  \bibinfo{journal}{Phys. Rev. D} \textbf{\bibinfo{volume}{84}},
  \bibinfo{pages}{094010} (\bibinfo{year}{2011}{\natexlab{a}}).

\bibitem[{\citenamefont{Aaron et~al.}(2010)}]{HERAcombined}
\bibinfo{author}{\bibfnamefont{F.~D.} \bibnamefont{Aaron}} \bibnamefont{et~al.}
  (\bibinfo{collaboration}{H1 and ZEUS}), \bibinfo{journal}{JHEP}
  \textbf{\bibinfo{volume}{1001}}, \bibinfo{pages}{109} (\bibinfo{year}{2010}),
  \eprint{arXiv:0911.0884 [hep-ex]}.

\bibitem[{\citenamefont{{Ashok \lowercase{s}uri }}(1971)}]{ashok}
\bibinfo{author}{\bibnamefont{{Ashok \lowercase{s}uri }}},
  \bibinfo{journal}{Phys. Rev. D} \textbf{\bibinfo{volume}{4}},
  \bibinfo{pages}{570} (\bibinfo{year}{1971}).

\bibitem[{\citenamefont{Martin}(1963)}]{martin1}
\bibinfo{author}{\bibfnamefont{A.}~\bibnamefont{Martin}},
  \bibinfo{journal}{Phys. Rev.} \textbf{\bibinfo{volume}{129}},
  \bibinfo{pages}{1432} (\bibinfo{year}{1963}).

\bibitem[{\citenamefont{Jin and Martin}(1964)}]{martin2}
\bibinfo{author}{\bibfnamefont{Y.~S.} \bibnamefont{Jin}} \bibnamefont{and}
  \bibinfo{author}{\bibfnamefont{A.}~\bibnamefont{Martin}},
  \bibinfo{journal}{Phys. Rev.} \textbf{\bibinfo{volume}{135}},
  \bibinfo{pages}{1375} (\bibinfo{year}{1964}).

\bibitem[{\citenamefont{Martin}(1966)}]{martin3}
\bibinfo{author}{\bibfnamefont{A.}~\bibnamefont{Martin}},
  \bibinfo{journal}{Nuovo Cimento} \textbf{\bibinfo{volume}{42}},
  \bibinfo{pages}{930} (\bibinfo{year}{1966}).

\bibitem[{\citenamefont{Froissart}(1961)}]{froissart}
\bibinfo{author}{\bibfnamefont{M.}~\bibnamefont{Froissart}},
  \bibinfo{journal}{Phys. Rev.} \textbf{\bibinfo{volume}{123}},
  \bibinfo{pages}{1053} (\bibinfo{year}{1961}).

\bibitem[{\citenamefont{Sakurai}(1960)}]{sakurai1}
\bibinfo{author}{\bibfnamefont{J.}~\bibnamefont{Sakurai}},
  \bibinfo{journal}{Ann. of Physics (NY)} \textbf{\bibinfo{volume}{11}},
  \bibinfo{pages}{1} (\bibinfo{year}{1960}).

\bibitem[{\citenamefont{Sakurai}(1969)}]{sakurai2}
\bibinfo{author}{\bibfnamefont{J.}~\bibnamefont{Sakurai}},
  \emph{\bibinfo{title}{Currents and Mesons}} (\bibinfo{publisher}{Univ. of
  Chicago Press}, \bibinfo{year}{1969}).

\bibitem[{\citenamefont{Schildknecht}(2006)}]{schildknecht}
\bibinfo{author}{\bibfnamefont{D.}~\bibnamefont{Schildknecht}},
  \bibinfo{journal}{Acta. Phys. Polon.} \textbf{\bibinfo{volume}{B37}},
  \bibinfo{pages}{595} (\bibinfo{year}{2006}).

\bibitem[{\citenamefont{Durand}(1985)}]{Snowmass84}
\bibinfo{author}{\bibfnamefont{L.}~\bibnamefont{Durand}}, in
  \emph{\bibinfo{booktitle}{Design and Utilization of the Superconducting
  Supercollider, Snowmass 1984}}, edited by
  \bibinfo{editor}{\bibfnamefont{P.}~\bibnamefont{Donaldson}} \bibnamefont{and}
  \bibinfo{editor}{\bibfnamefont{J.}~\bibnamefont{Morfin}}
  (\bibinfo{publisher}{Division of Particles and Fields of the American
  Physical Society}, \bibinfo{year}{1985}), p. \bibinfo{pages}{258}.

\bibitem[{\citenamefont{L'Heureux et~al.}(1985)\citenamefont{L'Heureux,
  Margolis, and Valin}}]{Margolis}
\bibinfo{author}{\bibfnamefont{P.}~\bibnamefont{L'Heureux}},
  \bibinfo{author}{\bibfnamefont{B.}~\bibnamefont{Margolis}}, \bibnamefont{and}
  \bibinfo{author}{\bibfnamefont{P.}~\bibnamefont{Valin}},
  \bibinfo{journal}{Phys. Rev. D} \textbf{\bibinfo{volume}{32}},
  \bibinfo{pages}{1681} (\bibinfo{year}{1985}).

\bibitem[{\citenamefont{Durand and Pi}(1987)}]{DurandPi}
\bibinfo{author}{\bibfnamefont{L.}~\bibnamefont{Durand}} \bibnamefont{and}
  \bibinfo{author}{\bibfnamefont{H.}~\bibnamefont{Pi}}, \bibinfo{journal}{Phys.
  Rev. Lett.} \textbf{\bibinfo{volume}{58}}, \bibinfo{pages}{303}
  (\bibinfo{year}{1987}).

\bibitem[{\citenamefont{Honjo et~al.}(1993)\citenamefont{Honjo, Durand, Gandhi,
  Pi, and Sarcevic}}]{HDGPS}
\bibinfo{author}{\bibfnamefont{K.}~\bibnamefont{Honjo}},
  \bibinfo{author}{\bibfnamefont{L.}~\bibnamefont{Durand}},
  \bibinfo{author}{\bibfnamefont{R.}~\bibnamefont{Gandhi}},
  \bibinfo{author}{\bibfnamefont{H.}~\bibnamefont{Pi}}, \bibnamefont{and}
  \bibinfo{author}{\bibfnamefont{I.}~\bibnamefont{Sarcevic}},
  \bibinfo{journal}{Phys. Rev. D} \textbf{\bibinfo{volume}{48}},
  \bibinfo{pages}{1048} (\bibinfo{year}{1993}).

\bibitem[{\citenamefont{Gribov et~al.}(1983)\citenamefont{Gribov, Levin, and
  Ryskin}}]{GLRsmxqcd}
\bibinfo{author}{\bibfnamefont{L.}~\bibnamefont{Gribov}},
  \bibinfo{author}{\bibfnamefont{E.}~\bibnamefont{Levin}}, \bibnamefont{and}
  \bibinfo{author}{\bibfnamefont{M.}~\bibnamefont{Ryskin}},
  \bibinfo{journal}{Phys. Reports} \textbf{\bibinfo{volume}{100}},
  \bibinfo{pages}{1} (\bibinfo{year}{1983}).

\bibitem[{\citenamefont{Gribov and Lipatov}(1972)}]{dglap1}
\bibinfo{author}{\bibfnamefont{V.~N.} \bibnamefont{Gribov}} \bibnamefont{and}
  \bibinfo{author}{\bibfnamefont{L.~N.} \bibnamefont{Lipatov}},
  \bibinfo{journal}{Sov. J. Nucl. Phys.} \textbf{\bibinfo{volume}{15}},
  \bibinfo{pages}{438} (\bibinfo{year}{1972}).

\bibitem[{\citenamefont{Altarelli and Parisi}(1977)}]{dglap2}
\bibinfo{author}{\bibfnamefont{G.}~\bibnamefont{Altarelli}} \bibnamefont{and}
  \bibinfo{author}{\bibfnamefont{G.}~\bibnamefont{Parisi}},
  \bibinfo{journal}{Nucl. Phys. B} \textbf{\bibinfo{volume}{126}},
  \bibinfo{pages}{298} (\bibinfo{year}{1977}).

\bibitem[{\citenamefont{Dokshitzer}(1977)}]{dglap3}
\bibinfo{author}{\bibfnamefont{Y.~L.} \bibnamefont{Dokshitzer}},
  \bibinfo{journal}{Sov. Phys. JETP} \textbf{\bibinfo{volume}{46}},
  \bibinfo{pages}{641} (\bibinfo{year}{1977}).

\bibitem[{\citenamefont{Breitweg et~al.}(2000)}]{ZEUS1}
\bibinfo{author}{\bibfnamefont{J.}~\bibnamefont{Breitweg}} \bibnamefont{et~al.}
  (\bibinfo{collaboration}{ZEUS Collaboration}), \bibinfo{journal}{Phys. Lett.
  B} \textbf{\bibinfo{volume}{487}}, \bibinfo{pages}{53}
  (\bibinfo{year}{2000}).

\bibitem[{\citenamefont{Chekanov et~al.}(2001)}]{ZEUS2}
\bibinfo{author}{\bibfnamefont{S.}~\bibnamefont{Chekanov}} \bibnamefont{et~al.}
  (\bibinfo{collaboration}{ZEUS Collaboration}), \bibinfo{journal}{Eur. Phys.
  J. C} \textbf{\bibinfo{volume}{21}}, \bibinfo{pages}{443}
  (\bibinfo{year}{2001}).

\bibitem[{\citenamefont{Adloff et~al.}(2001)}]{H1}
\bibinfo{author}{\bibfnamefont{C.}~\bibnamefont{Adloff}} \bibnamefont{et~al.}
  (\bibinfo{collaboration}{H1 Collaboration}), \bibinfo{journal}{Eur. Phys. J.
  C} \textbf{\bibinfo{volume}{21}}, \bibinfo{pages}{33} (\bibinfo{year}{2001}).

\bibitem[{\citenamefont{Block}(2006{\natexlab{b}})}]{sieve}
\bibinfo{author}{\bibfnamefont{M.~M.} \bibnamefont{Block}},
  \bibinfo{journal}{Nucl. Inst. and Meth. A.} \textbf{\bibinfo{volume}{556}},
  \bibinfo{pages}{308} (\bibinfo{year}{2006}{\natexlab{b}}).

\bibitem[{\citenamefont{Ellis et~al.}(2003)\citenamefont{Ellis, Stirling, and
  Webber}}]{esw}
\bibinfo{author}{\bibfnamefont{R.~K.} \bibnamefont{Ellis}},
  \bibinfo{author}{\bibfnamefont{W.~J.} \bibnamefont{Stirling}},
  \bibnamefont{and} \bibinfo{author}{\bibfnamefont{B.~R.}
  \bibnamefont{Webber}}, \emph{\bibinfo{title}{{QCD} and {Collider Physics}}}
  (\bibinfo{publisher}{Cambridge University Press}, \bibinfo{year}{2003}).

\bibitem[{\citenamefont{Bardeen et~al.}(1978)\citenamefont{Bardeen, Buras,
  Duke, and Muta}}]{bardeen}
\bibinfo{author}{\bibfnamefont{W.~A.} \bibnamefont{Bardeen}},
  \bibinfo{author}{\bibfnamefont{A.~J.} \bibnamefont{Buras}},
  \bibinfo{author}{\bibfnamefont{D.~W.} \bibnamefont{Duke}}, \bibnamefont{and}
  \bibinfo{author}{\bibfnamefont{T.}~\bibnamefont{Muta}},
  \bibinfo{journal}{Phys. Rev. D} \textbf{\bibinfo{volume}{18}},
  \bibinfo{pages}{3998} (\bibinfo{year}{1978}).

\bibitem[{\citenamefont{Harrod and Wada}(1980)}]{hw}
\bibinfo{author}{\bibfnamefont{R.}~\bibnamefont{Harrod}} \bibnamefont{and}
  \bibinfo{author}{\bibfnamefont{S.}~\bibnamefont{Wada}},
  \bibinfo{journal}{Phys. Lett.} \textbf{\bibinfo{volume}{96B}},
  \bibinfo{pages}{195} (\bibinfo{year}{1980}).

\bibitem[{\citenamefont{Furmanski and Petronzio}(1982)}]{furmanski}
\bibinfo{author}{\bibfnamefont{W.}~\bibnamefont{Furmanski}} \bibnamefont{and}
  \bibinfo{author}{\bibfnamefont{R.}~\bibnamefont{Petronzio}},
  \bibinfo{journal}{Zeit. fur Physik} \textbf{\bibinfo{volume}{C11}},
  \bibinfo{pages}{293} (\bibinfo{year}{1982}).

\bibitem[{\citenamefont{Block et~al.}(2008)\citenamefont{Block, Durand, and
  McKay}}]{bdm1}
\bibinfo{author}{\bibfnamefont{M.~M.} \bibnamefont{Block}},
  \bibinfo{author}{\bibfnamefont{L.}~\bibnamefont{Durand}}, \bibnamefont{and}
  \bibinfo{author}{\bibfnamefont{D.~W.} \bibnamefont{McKay}},
  \bibinfo{journal}{Phys. Rev. D} \textbf{\bibinfo{volume}{77}},
  \bibinfo{pages}{094003} (\bibinfo{year}{2008}), \eprint{arXiv:0710.3212
  [hep-ph]}.

\bibitem[{\citenamefont{Block et~al.}(2009)\citenamefont{Block, Durand, and
  McKay}}]{bdm2}
\bibinfo{author}{\bibfnamefont{M.~M.} \bibnamefont{Block}},
  \bibinfo{author}{\bibfnamefont{L.}~\bibnamefont{Durand}}, \bibnamefont{and}
  \bibinfo{author}{\bibfnamefont{D.~W.} \bibnamefont{McKay}},
  \bibinfo{journal}{Phys. Rev. D} \textbf{\bibinfo{volume}{79}},
  \bibinfo{pages}{014031} (\bibinfo{year}{2009}), \eprint{arXiv:0808.0201
  [hep-ph]}.

\bibitem[{\citenamefont{Block et~al.}(2010{\natexlab{b}})\citenamefont{Block,
  Durand, Ha, and McKay}}]{bdhmNLO}
\bibinfo{author}{\bibfnamefont{M.~M.} \bibnamefont{Block}},
  \bibinfo{author}{\bibfnamefont{L.}~\bibnamefont{Durand}},
  \bibinfo{author}{\bibfnamefont{P.}~\bibnamefont{Ha}}, \bibnamefont{and}
  \bibinfo{author}{\bibfnamefont{D.~W.} \bibnamefont{McKay}},
  \bibinfo{journal}{Eur. Phys. J. C} \textbf{\bibinfo{volume}{69}},
  \bibinfo{pages}{425} (\bibinfo{year}{2010}{\natexlab{b}}),
  \eprint{arXiv:1005.2556 [hep-ph]}.

\bibitem[{\citenamefont{Lai et~al.}(2010)\citenamefont{Lai, Guzzi, Huston, Li,
  Nadolsky, Pumplin, and Yuan}}]{CT10}
\bibinfo{author}{\bibfnamefont{H.-L.} \bibnamefont{Lai}},
  \bibinfo{author}{\bibfnamefont{M.}~\bibnamefont{Guzzi}},
  \bibinfo{author}{\bibfnamefont{J.}~\bibnamefont{Huston}},
  \bibinfo{author}{\bibfnamefont{Z.}~\bibnamefont{Li}},
  \bibinfo{author}{\bibfnamefont{P.~M.} \bibnamefont{Nadolsky}},
  \bibinfo{author}{\bibfnamefont{J.}~\bibnamefont{Pumplin}}, \bibnamefont{and}
  \bibinfo{author}{\bibfnamefont{C.-P.} \bibnamefont{Yuan}},
  \bibinfo{journal}{Phys. Rev. D} \textbf{\bibinfo{volume}{82}},
  \bibinfo{pages}{072024} (\bibinfo{year}{2010}),
  \eprint{\lowercase{a}rXiv:1007.2241[hep-ph]}.

\bibitem[{Dur()}]{Durham}
\bibinfo{note}{\lowercase{h}ttp://durpdg.dur.ac.uk/hepdata/pdf3.html}.

\bibitem[{\citenamefont{Block et~al.}(2011{\natexlab{b}})\citenamefont{Block,
  Durand, Ha, and McKay}}]{bdhmLO}
\bibinfo{author}{\bibfnamefont{M.~M.} \bibnamefont{Block}},
  \bibinfo{author}{\bibfnamefont{L.}~\bibnamefont{Durand}},
  \bibinfo{author}{\bibfnamefont{P.}~\bibnamefont{Ha}}, \bibnamefont{and}
  \bibinfo{author}{\bibfnamefont{D.~W.} \bibnamefont{McKay}},
  \bibinfo{journal}{Phys. Rev. D} \textbf{\bibinfo{volume}{{\bf 83}}},
  \bibinfo{pages}{054009} (\bibinfo{year}{2011}{\natexlab{b}}),
  \eprint{arXiv:1010.2486 [hep-ph]}.

\end{thebibliography}

\end{document}